\documentclass[11pt,draftcls,onecolumn]{IEEEtran}
\usepackage{graphicx}
\usepackage{cite}
\usepackage{array}
\usepackage{tabulary}
\newcolumntype{K}[1]{>{\centering\arraybackslash}p{#1}}

\usepackage{chemarrow}
\usepackage{color}
\usepackage{algorithmic}
\usepackage{algorithm}
\usepackage{clrscode}
\usepackage{amsmath, amssymb, mathrsfs, amsfonts, amsthm}
\usepackage{epsfig, epstopdf}
\usepackage{multirow}

\allowbreak
\usepackage{caption}
\usepackage{stfloats}
\usepackage{subfig}
\usepackage{enumerate}
\usepackage{float}

\theoremstyle{theorem}
\newtheorem{lemma}{Lemma}

\newtheorem{remark}{Remark}

\def\msT{ \mathsf{T}}
\def\mP{\mathbb{P}}

\def\msR{\mathsf{R}}
\def\msM{\mathsf{M}}
\def \msA{\mathsf{A}}
\def \msB{\mathsf{B}}
\def \msC{\mathsf{C}}
\def \msD{\mathsf{D}}

\newcommand\blfootnote[1]{%
  \begingroup
  \renewcommand\thefootnote{}\footnote{#1}%
  \addtocounter{footnote}{-1}%
  \endgroup
}

\allowdisplaybreaks
\date{}
\title{On Medium Chemical Reaction in Diffusion-Based Molecular Communication: a Two-Way Relaying Example}
\author{\IEEEauthorblockN{Maryam Farahnak-Ghazani, Gholamali Aminian, Mahtab Mirmohseni, Amin Gohari, and Masoumeh Nasiri-Kenari}\\
\IEEEauthorblockA{Sharif University of Technology}}

\begin{document}
\maketitle
\begin{abstract}
Chemical reactions are a prominent feature of molecular communication (MC) systems, with no direct parallels in wireless communications. While chemical reactions may be used inside the transmitter nodes, receiver nodes or the communication medium, we focus on its utility in the medium in this paper. Such chemical reactions can be used to perform computation over the medium as molecules diffuse and react with each other (physical-layer computation). We propose the use of chemical reactions for the following purposes: (i) to reduce signal-dependent observation noise of receivers by reducing the signal density, (ii) to realize molecular physical-layer network coding (molecular PNC) by performing the natural XOR operation inside the medium, and (iii) to reduce the inter-symbol interference (ISI) of other transmitters by canceling out the remaining molecules from previous transmissions. To make the ideas formal, we consider an explicit two-way relaying example with a transparent receiver (which has a signal-dependent noise). The proposed ideas are used to define a modulation scheme (which we call the PNC scheme). We compare the PNC with a previously proposed scheme for this problem where the XOR operation is performed at the relay node (using a molecular logic gate). We call the latter, the straightforward network coding (SNC).  It is observed that in addition to the simplicity of the proposed PNC scheme, it outperforms the SNC scheme especially when we consider ISI.\blfootnote{This work was in part presented in the 2016 Iran Workshop on Communication and Information Theory (IWCIT) \cite{IWCIT2016physical}.}

\end{abstract}

\section{Introduction}
While traditional wireless communication systems employ \emph{energy carriers} (such as electromagnetic or acoustic waves) for communication, Molecular Communication (MC) utilizes \emph{physical molecules} as its carriers of information. In {diffusion-based} MC system, the transmitter and the receiver are biological/engineered cells or electronic systems that release or receive molecules, while the channel is assumed to be a fluid medium in which molecules diffuse.
Electromagnetic waves and molecular diffusion share similarities and differences.
Both the electromagnetic wave equation and the Fick's second law of macroscopic diffusion are second-order linear partial differential equations. As a result, both lead to linear system models that satisfy the superposition property. 
%The superposition property is used in the design of multi--carrier wireless systems. 
However, there are also differences between electromagnetic waves and molecular diffusion. Notably, the degradation and attenuation of transmitted signals are more pronounced in molecular diffusion-based channels and seriously limit the transmission distance between the transmitter and the receiver \cite{fekrirelay1}. Relaying is a solution for increasing the range of communication and has been utilized by nature in intracellular communication \cite[Chapter~15]{Alberts2002}. In addition, while the measurement noise of a wireless receiver may be modeled by an additive Gaussian noise (the AWGN channel), some of the most promising molecular receptors, such as the ligand receiver and the transparent receiver, have a signal-dependent measurement noise (\emph{i.e.,} their noise variance is higher when they are measuring a larger signal) \cite{gohari2016information, Pierobon2011, aminian2016}.
Furthermore, when there are multiple molecule types in the medium, they may undergo chemical reactions as they diffuse in the environment. The reaction amongst the molecules is governed by the non-linear reaction--diffusion differential equations.

Chemical reaction is a key operation mechanism of biological systems. As a result, chemical reactions are likely to be a fixture of future engineered molecular transmitters or receivers. For instance, \cite{chou2015markovian, chou2015impact} consider the role of chemical reaction in transmitter and receiver design. However, the emphasis of this paper is on the challenges and opportunities of utilizing chemical reactions \emph{inside the communication medium (channel)} rather than inside the transmitter or receiver nodes. We may view the diffusion-reaction process as a form of \emph{physical-layer computation} that is performed over the medium (distinct from the operation of transceiver cells). While the superposition property has been utilized for ``computation over the air" in the wireless literature \cite{Nazer, Goldenbaum, Limmer, Abari, liew2013physical}, chemical reactions provide the possibility of more complicated interactions than a simple superposition. Although few existing works provide a number of ideas for exploiting chemical reactions in the medium for communication purposes, we still lack a full understanding.  In this paper, we review the state of the art and give a number of new ideas. In particular, our emphasis is on the utility of chemical reactions by the relay nodes.

\textbf{Challenges and known techniques:}
While linear chemical reactions can be readily utilized for signal shaping, the more interesting chemical reactions are non-linear and demonstrate complicated patterns \cite{samoilov2002signal}.
The main challenge of utilizing chemical reactions is the non-linearity of the reaction-diffusion equations and lack of explicit analytical solutions. %of the simplest reaction-diffusion equations.
For instance, consider the following chemical reaction: %between two molecule types of $\msA$ and $\msB$:
\begin{align}\label{eqn:chemical-reaction}
\msA+\msB \underset{\kappa}{\overset {\gamma}{\rightleftharpoons}} \msC
\end{align}
in which $\gamma$ and $\kappa$ are the forward and reverse reaction rate constants, respectively. Let $c_\msA$, $c_\msB$, and $c_\msC$ be the concentrations of $\msA$, $\msB$, and $\msC$, respectively. The reaction-diffusion law can be expressed as \cite{Kuttler2011}
\begin{equation}\label{eqav}
\begin{aligned}
\frac{\partial c_\msA}{\partial t} = D_\msA \nabla ^2 c_\msA
-\gamma c_\msA c_\msB+\kappa c_{\msC}, \quad
\frac{\partial c_\msB}{\partial t} = D_\msB \nabla ^2 c_\msB
-\gamma c_\msA c_\msB+\kappa c_{\msC}, \quad
\frac{\partial c_\msC}{\partial t} = D_\msC \nabla ^2 c_\msC
+\gamma c_\msA c_\msB-\kappa c_{\msC},
\end{aligned}
\end{equation}
where $D_\msA$, $D_\msB$, and $D_{\msC}$ are the diffusion coefficients of $\msA$, $\msB$, and $\msC$, respectively.
The term $\gamma c_\msA c_\msB$ is the challenging non-linear term. Thus far, this challenge is mostly dealt with in the MC literature by noting that despite lack of analytical solutions, it may be still possible to intuitively predict the \emph{qualitative behavior} of the solutions, %of the reaction--diffusion systems,
in particular when the reaction is limited to a small neighborhood \cite{Gold} or is instantaneous (high forward reaction rate constant and low reverse reaction rate constant). The general approach is to use the high-level intuition to design signaling schemes, which may be backed up with numerical simulations or partial supporting analysis.
%We give a number of new ideas for going beyond this general approach in Section \ref{sec:futurework}.

We may categorize the known ideas of utilizing chemical reactions in the medium as follows:
\begin{itemize}
\item \emph{Memory degradation:}  In \cite{enzym1}, it is suggested to release enzymes throughout the environment.\footnote{While  \cite{enzym1} assumes enzymes are released throughout the medium, \cite{cho2017effective} studies its release in a limited area of the medium.} A chemical reaction between enzymes and information carrying molecules cancels out the involved molecules, and has the effect of shortening the lifetime distribution of all molecules in the environment. This reaction can put down inter-symbol interference (ISI) by reducing the remaining molecular concentration from previous transmissions, at the cost of weakening the desired signal.% direct link between the transmitter and the receiver.

\item \emph{Pattern formation:} In the above item, we gave a chemical reaction that simply reduces the concentration of the reactant molecules. However, more complicated dynamics and patterns (such as oscillating reactions or traveling waves) can arise from chemical reactions. Assuming that molecules of type $\msA$ are used for communication, it has been suggested in \cite{PatternFormation} to fill the environment with molecules of type $\msB$ whose reaction with molecules of type $\msA$ produces such oscillating and propagating patterns. This may be utilized to increase the propagation range of the molecules (before they dissolve in the environment). The more complicated spatial-temporal patterns could increase the decoder's ability to distinguish amongst them; this can effectively increase the information capacity of the system.

\item \emph{Simulating negative signals and ISI reduction:} Unlike electrical current and voltage that can take negative values, the density of molecules in an environment cannot go negative. %as only a positive concentration of molecules may be released in the medium.
Chemical reactions are proposed for simulating transmission of a negative signal by a molecular transmitter \cite{newadd1, Gold, IWCIT2016type}. For instance, authors of \cite{Gold} suggest using $\text{H}^{+}$ and $\text{OH}^{-}$ ions. Release of any of these ions reduces the concentration of the other one in the medium, and one can interpret release of $\text{H}^{+}$ ions as sending a positive, and release of $\text{OH}^{-}$ ions as sending a negative signal. Further, Simulation of negative signals allows for design of precoders at the transmitter to combat the ISI (e.g. see \cite{IWCIT2016type}).

\item \emph{Relay signal amplification:} Authors in \cite{nakano2011repeater} describe a chemical reaction that amplifies the incoming signals. However, we point out that signal amplification may be also performed blindly in the medium; assume that the information molecule is of type $\msA$ and the relay releases a limited number of molecules of type $\msB$ such that
\begin{align}\label{eqn:chemical-reaction23}
\msA+\msB \underset{\kappa}{\overset {\gamma}{\rightleftharpoons}} 2\msC+\msD.
\end{align}
This reaction produces molecules of type $\msC$ whose concentration is twice the concentration of molecules of type $\msA$ in the environment. Thus, the relay simply releases molecules of type $\msB$ without having to sense the incoming density of molecules of type $\msA$.
\item\emph{Molecular media-based modulation}: Authors in \cite{gohari2016information} argue that information can be transmitted by changing the general physical properties of the communication medium (rather than directly changing the density of the released molecules). For instance, assume that we have two transmitters, called the $\msA$-transmitter and the $\msB$-transmitter, which release molecules of types $\msA$ and $\msB$ in the medium, respectively. There is a receiver which can \emph{only} sense the density of molecules of type $\msA$. If $\msA$ and $\msB$ react in the environment, the $\msB$-transmitter can communicate indirectly to the receiver (despite the receiver only has sensors that detect $\msA$ molecules): the reason is that the actions of the $\msB$-transmitter influences the communication medium between the $\msA$-transmitter and the receiver.
\end{itemize}

Besides the above explicit ideas for medium chemical reactions, authors in \cite{karig2011model} utilize an interesting feature of non-linear systems, namely harnessing noise for signal propagation in a cell-to-cell MC system. Unlike linear systems where noise plays a disruptive role, noise can increase information capacity of non-linear systems (this effect is known as the \emph{stochastic resonance}).

\textbf{Our contribution:}
Our main contribution in this work is to propose new ideas for utility of chemical reactions in a communication medium. These ideas are as follows:
\begin{enumerate}
\item \emph{Receiver noise reduction:} As mentioned earlier, many molecular receivers have signal dependent noise. In particular, they face a smaller noise if they are sensing a smaller signal. Now, suppose the density of molecules around the receiver is $y$ and the receiver wants to measure it. If a receiver can predict that $y$ is at least $\lambda$, it can locally release a different species of molecules that would react with the signal molecules around the receiver, and reduce the signal molecule density by $\lambda$ in the vicinity of the receiver. Thus, instead of measuring $y$, it measures $y-\lambda$. This will incur a smaller signal dependent noise.
The receiver can predict a minimum value for its upcoming measurement $y$ by utilizing its previous observations. For instance, if the receiver has measured a high density of molecules in the previous time slot, it expects the density of molecules to be high in the current time slot as well. The reason is that diffusion is a slow process and it takes time for the effect of previous transmissions to disappear from the medium.
%As a result, the receiver may have an estimate that the molecule density is at least $\lambda$, where $\lambda$ is found adaptively from its previous observations.
One should also consider the possibility that the estimate $\lambda$ is incorrect, \emph{i.e.,} $y$ is less than $\lambda$. In this case, the receiver observes $\min(0, y-\lambda)=0$, and the information about $y$ will be lost. Receiver's error in finding a suitable lower bound $\lambda$ for $y$ can result in an error, but the probability of this error can be small and compensate for the decrease in the signal-dependent measurement noise.\footnote{We have already used a simpler form of this idea in \cite{IWCIT2016type}, but in that work the amount of release of molecules was not chosen adaptively by the receiver.}

\item \emph{Molecular physical-layer network coding (Molecular PNC):} In traditional wireless networks, due to the broadcast nature, network coding can be used by the nodes to improve the throughput of the system. Two main classes of network coding schemes in traditional wireless networks are straightforward network coding (SNC) and physical layer network coding (PNC) \cite{networkcoding}. SNC in MC has been studied in \cite{akan2013, aghvami2014}, where the relay uses an XOR logic gate \cite{credi}, at the molecular level, to XOR the messages of the two transceivers. The traditional PNC is based on the fact that the signals can become negative and thus may cancel out each other physically when they superpose in the environment. Since in MC the transmitted signals cannot become negative, we suggest the use of molecular reaction to cancel out the signals and realize the XOR operation inside the medium. This allows for removal of the XOR gate inside the relay node.\footnote{The authors in \cite{Oishi2011} discuss implementing general linear systems with chemical reactions (more broadly than an XOR like operation).}
%As we show later, one can improve upon the previously proposed schemes by realizing the XOR operation inside the medium via chemical reactions. This allows for removal of the XOR gate inside the relay node.
The idea is as follows: suppose we have molecules of type $\msA$ and $\msB$ that react and cancel out each other. Then, if only one molecule type exists in the medium, it survives. However, the presence of both molecules results in the destruction of both.

\item \emph{The dual purpose of transmission:} Thus far, the literature assumes that a transmitter releases molecules to convey its own message. Consider a scenario where we have two nodes that are using molecules of types $\msA$ and $\msB$ for transmission, respectively. These transmitters also have receptors that allows them to obtain information about the other node's transmissions.
Assume that these molecules of types $\msA$ and $\msB$ can react and cancel out each other. Then, the first node can release molecules of type $\msA$ for (i) encoding of its information bits, or (ii) for reducing the density of the other node's molecule to reduce its measurement noise level.
\end{enumerate}

\textbf{Example of a two-way relay network model:} To make the above ideas formal at once, we propose a specific setup with a certain signal-dependent receiver noise. We give an explicit modulation scheme that utilizes all the above-mentioned ideas in its design. More specifically, we consider a two-way molecular relay network, where two nano-transceivers exchange their information through a nano-relay. For this network, in this paper, we propose a new network coding scheme in MC parallel to the PNC in traditional wireless networks. This covers our two new ideas (namely receiver noise reduction and molecular PNC) mentioned above.
%By making physical-layer XOR using reaction, the signal density reduces when both molecules arrive at the relay and hence the signal dependent noise at the relay is reduced.
We show that our proposed PNC scheme outperforms the previously proposed SNC scheme for MC.

A complication arises if the above molecular channels have ISI, and this is where we make use of our third new idea (the dual purpose of transmission). For point-to-point channels, ISI mitigating techniques have been introduced in \cite{mosayebi2014},\cite{movahednasab2015}. However, to the best knowledge, there is no study on the ISI-mitigating schemes in two-way relay channels. One natural way to tackle this problem is to apply the point-to-point ISI mitigating techniques to each hop of the relay channel. For the SNC scheme, we extend the existing ISI mitigating techniques of point-to-point channels proposed in \cite{mosayebi2014}, \cite{movahednasab2015} to each hop. However, for the PNC scheme we propose a novel ISI-mitigating scheme, which is based on two observations: i) in two-way channels each transceiver has access to the previous messages of the other transceiver, and thus knows an estimation of the other user's ISI. ii) The molecular reaction can be used to cancel out the ISI (or reduce the estimated ISI). 

% It is important to point out that in a channel with ISI, a transmitter may release molecules even when its bit is zero; this is to cancel out the ISI of the other receiver.

We make the following conclusions from our analysis of the proposed molecular PNC scheme. In the no ISI case, our results (based on the derived closed form equations for the transparent receiver) show that the PNC outperforms the SNC in terms of error probability thanks to the reaction among the molecules in the PNC scheme. In fact, when the messages of both transceivers are 1, the number of the counted molecules at the receptors is reduced compared to the SNC scheme. This results in less error caused by the transparent receiver. These results are confirmed by simulations. In presence of ISI, the error probability of both ISI-mitigated PNC and SNC schemes are derived analytically (and confirmed by simulation); it is shown that the PNC performs significantly better than the SNC.

This paper is organized as follows: in Section \ref{sec:model}, we present the physical model for the two-way relay example. In Section \ref{sec:pncscheme}, we describe the use of chemical reaction for molecular PNC and receiver noise reduction, and in Section \ref{sec:ISImitigate}, we explain the idea of chemical reaction for dual purpose of transmission. In Section \ref{error} and \ref{errorISI}, the error performance of the two schemes in no ISI and ISI cases are respectively investigated.
%and in Section \ref{errorISI}, we investigate the error performance of the schemes in the presence of ISI using ISI mitigating techniques.
In Section \ref{Simulation}, we present the numerical results, and finally, we include concluding remarks in Section \ref{conclusion}.

\textbf{Notation:}
%We use the superscript $\msT_i\msR$ for the parameters of the channel from the transceiver $\msT_i$ to the relay, $\msR\msT_i$ for the parameters of the channel from the relay to the transceiver $\msT_i$.
The random variables, error events, and diffusion coefficients are shown by upper cases while the realizations of random variables are indicated by lower cases. The event $E^c$ shows the complement of the event $E$ and $\bar{i}$ denotes the complement of $i$ in its defined set. The decoded value of the information bit $B$ is denoted by $\hat{B}$.

%\vspace{-0.5em}
\section{Physical Model }\label{sec:model}
We consider a diffusion-based nano-network consisting of two nano-transceivers and a nano-relay with the ability of both transmitting and receiving information in different time slots. A two-way communication between two nano-transceivers is established by a nano-relay.
%The transceiver $\msT_1$, $\msT_2$ and the relay $\msR$ are located at points $r_1$, $r_2$, and $r_3$, respectively. 
%The distance of the relay from the transceiver $\msT_i$ is denoted as $d_i$.
For simplicity of the notations, we assume that the relay is in the same distance $d$ from the two transceivers. The transceiver $\msT_i$ for $i=1,2$ has a sequence of information bits $(B_{i,1}, B_{i,2}, \cdots)$ that wants to transmit to the other transceiver.

We assume that the time is slotted with duration $t_s$, and during any communication protocol, molecules are released by either the transceiver $\msT_i$ or relay $\msR$ at the beginning of the time slots. For instance, a protocol might utilize the on-off keying (OOK) modulation for transmission in which each transmitter releases a burst of molecules to send the information bit $1$ at the beginning of each time slot, or stays silent to send the information bit $0$. We assume that $\msT_1$ releases molecules of type $\msM_1$, $\msT_2$ releases molecules of type $\msM_2$, and the relay releases molecule type $\msM_3$ (to avoid self-interference \cite{schober1}).
The diffusion coefficients of molecules of type $\msM_i$ are noted by $D_i$.
Again for simplicity, we assume $D_i=D$ for $i=1,2,3$. While molecules are released at the beginning of time slots of duration $t_s$, molecule density is measured by the receptors of $\msT_1$, $\msT_2$ or $\msR$ at time instances $t_0, t_0+t_s, t_0+2t_s, \dots$ for some $t_0 \leq t_s$.

\textbf{Channel model:} For the diffusion of molecules, we use the deterministic model based on Fick's second law of diffusion. According to this model, when there is no reaction among molecules of different types, the concentration of molecules of type $\msM_i$ at point $\vec{r}$ and time $t$, $c_i(\vec{r},t)$, is the solution of the following differential equation
\begin{equation}\label{eqav2}
\begin{aligned}
\frac{\partial c_i(\vec{r},t)}{\partial t} = D_i \nabla ^2 c_i(\vec{r},t)+g_i(\vec{r},t),
\end{aligned}
\end{equation}
where $g_i(\vec{r},t)$ is the concentration of released molecules of type $\msM_i$ at point $\vec{r}$ and time $t$. Since we assume the same diffusion coefficients for all molecule types, the impulse responses of the channels
(which are obtained when $g_i(\vec{r},t)=\delta(\vec{r})\delta(t)$)
 are the same, and for 3-D diffusion is obtained as \cite{Pierobon1}
\begin{equation}
\begin{aligned}\label{impulseresp}
h(\vec{r},t)=\frac{1[t>0]}{(4 \pi D t)^{{3}/{2}}}e^{-\frac{\|\vec{r}\|^2}{4 D t}}.
\end{aligned}
\end{equation}
Since the system is linear and time invariant (LTI), we have $c_i(\vec{r},t)=g_i(\vec{r},t)*h(\vec{r},t)$. This means that when a nano-transmitter, located at the origin, releases $\zeta_i$ molecules at time $t=0$, the concentration of molecules at point $\vec{r}$ and time $t$ will be $c_i(\vec{r},t)=\zeta_i h(\vec{r},t)$.\footnote{Note that if the transmitter of molecule type $\msM_i$ is located at $\vec{r}_i$, we should find the response of the system for input $g_i(\vec{r},t)=\delta(\vec{r}-\vec{r}_i)\delta(t)$, which will be $c_i(\vec{r},t)=\zeta_i h(\vec{r}-\vec{r}_i,t)$. However, as only the distance of the transmitter to the receiver appears in $h(\vec{r},t)$, for simplicity, we obtain the concentration of each molecule type assuming that its transmitter is located at the origin.} 

\textbf{Reception model:} Molecules released by $\msT_1$ and $\msT_2$ need to be measured by the relay $\msR$, and molecules released by the relay $\msR$ need to be measured by $\msT_1$ and $\msT_2$. Hence, we require two receptors for molecules of types $\msM_1$ and $\msM_2$ at the relay and one receptor to receive molecule type $\msM_3$ at each transceiver. The receivers at the transceivers and the relay are assumed to be \emph{transparent receivers}, in which the receiver counts the number of molecules that enter within its counting volume $v_\textrm{r}$ perfectly. We assume that the radius of the receiver is small with respect to the distance of the relay from the transceivers, and hence the concentration of molecules is almost uniform in this volume. For simplicity, we assume the same counting volume $v_\textrm{r}$ for the receivers at the transceivers and the relay. According to this model, when the average concentration of molecules of type $\msM_i$ at the receiver at time $t$ is $c_i(\vec{r},t)$, the average number of counted molecules in a volume $v_\textrm{r}$ is $c_i(\vec{r},t)v_\textrm{r}$, and the total number of counted molecules of type $\msM_i$ at time $t$ will follow a Poisson distribution with parameter $c_i(\vec{r},t)v_\textrm{r}$; see \cite{gohari2016information, Pieroborn2011transparent}.\footnote{The results of the paper can be easily extended for the general case with different distances from the transceivers to the relay, diffusion coefficients and receiver volumes.}

%\vspace{-0.5em}
\section{Chemical Reaction for Molecular PNC and Receiver Noise Reduction} \label{sec:pncscheme}
Here, we demonstrate the benefit of chemical reaction for molecular PNC and receiver noise reduction (as discussed in the introduction) in the context of the above two-way communication channel. We first describe the SNC scheme in part A. We then describe our proposed PNC scheme in part B.
In this section, we consider a channel with no ISI. The case with ISI is considered in Section~\ref{sec:ISImitigate} to illustrate the idea of the dual purpose of transmission.

%\vspace{-0.5em}
\subsection{The Previously Known SNC Scheme}
For the transmission model, we restrict to protocols in which the transceivers and the relay alternate in becoming active. In other words, in each run of the protocol, the transceivers $\msT_1$ and $\msT_2$ first become active and transmit molecules. Then, $\msT_1$ and $\msT_2$ become silent and the relay $\msR$ starts transmitting. During the $k$-th run of this protocol, %the transceiver
$\msT_i$ aims to communicate the bit $B_{i,k}$ to the other transceiver for $i=1,2$. This protocol is run repeatedly so that %the transceiver
$\msT_1$ reconstructs $(\hat B_{2,1}^{\msT_1}, \hat B_{2,2}^{\msT_1}, \cdots)$ while
% the transceiver
 $\msT_2$ reconstructs $(\hat B_{1,1}^{\msT_2}, \hat B_{1,2}^{\msT_2}, \cdots)$.
Since $\msT_1$ and $\msT_2$ use different molecule types, the transmission protocol needs two time slots in total. We consider a super time slot which contains two time slots of equal duration of $t_s$. Throughout the paper, $k$ shows the index of the super time slot. The communication protocol in this scheme has two phases described as follows (see Fig. \ref{NCmodel}):
\begin{figure}
\centering
\includegraphics[scale=0.28]{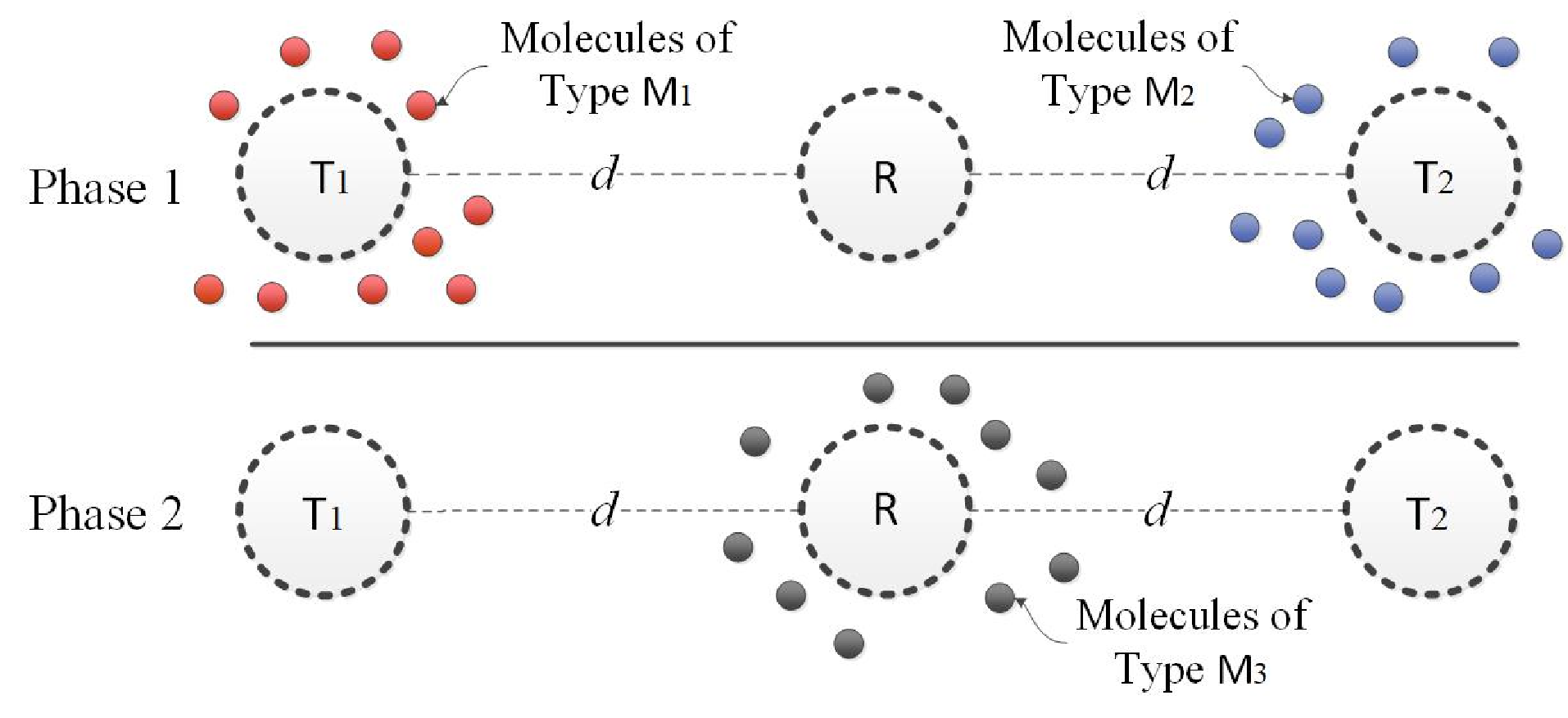}
\caption{A molecular two-way relay network}
\label{NCmodel}
\vspace{-1.5em}
\end{figure}
\begin{itemize}
\item \textbf{Phase 1:} In the first phase, the transceivers, $\msT_1$ and $\msT_2$, send their information bits, $B_{1,k}$ and $B_{2,k}$, at the beginning of the $k$-th super time slot to the relay using OOK modulation. Transceivers use different molecule types and this phase takes only one time slot. Employing the OOK modulation, the transceiver $\msT_i$ release $X_{i,k}=B_{i,k} \zeta_i$ molecules of types $\msM_i$, where $\zeta_i$ shows the number of the released molecules.

\item\textbf{Phase 2:} In the second phase, the relay decodes the messages of $\msT_1$ and $\msT_2$ as $\hat{B}_{1,k}^\msR$ and $\hat{B}_{2,k}^\msR$, respectively, and transmits the XOR of the decoded bits, $B_{\msR,k}=\hat{B}_{1,k}^\msR \oplus \hat{B}_{2,k}^\msR$, to both transceivers using OOK modulation, i.e., the relay releases
$X_{3,k}=B_{\msR,k} \zeta_3$ molecules of types $\msM_3$ in the $k$-th super time slot.
\end{itemize}
Each transceiver $\msT_i$ decodes the message of the relay as $\hat{B}_{\msR,k}^{\msT_i}$ and, by XORing it with its own sent message, finds the message sent by the other transceiver, i.e., $\hat{B}_{\bar{i},k}^{\msT_i}=B_{i,k} \oplus \hat{B}_{\msR,k}^{\msT_i}, i \in\{1,2\}$. The block diagram of the system is shown in Fig. \ref{blockdiag}.
\begin{figure}
\centering
\includegraphics[scale=0.5]{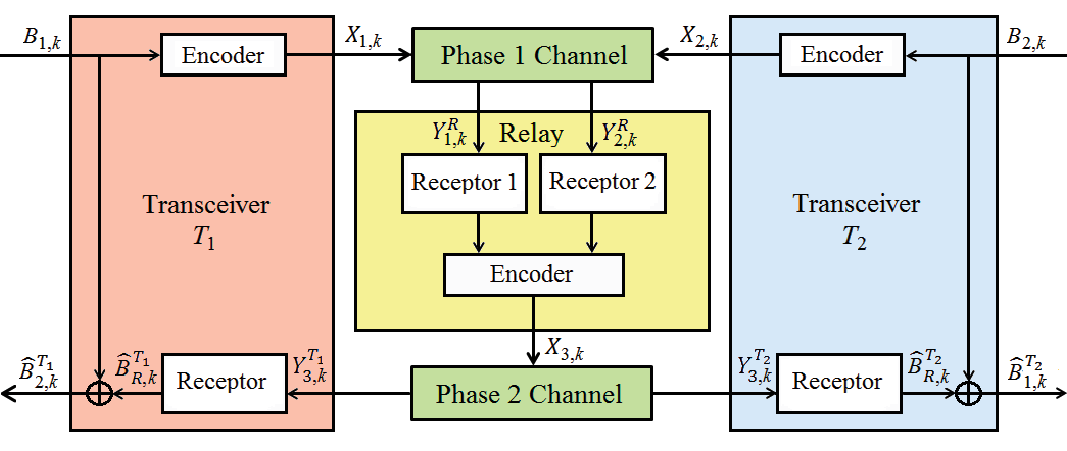}
\caption{Block diagram of the system in the SNC or PNC scheme}
\label{blockdiag}
\vspace{-1.5em}
\end{figure}

\textbf{Received concentration:} When we have no ISI in the channels, the remaining molecules of the previous super time slots are cleared from the medium before new molecules are released, and hence the average concentration of the molecules of type $\msM_i$ measured by its receptor type in the $k$-th super time slot is
\begin{equation}\label{eqCiknoISI}
\begin{aligned}
C_{i,k}=c_i(\vec{r},t_0)\big\rvert_{\|\vec{r}\|=d}=X_{i,k} \pi_1=B_{i,k}\zeta_i\pi_1, \qquad i \in \{1,2,3\}.
\end{aligned}
\end{equation}
where $\pi_l$ is the channel gain in the $l$-th time slot, which is obtained by sampling $h(\vec{r},t)$ at time instance $t_0+(l-1)t_s$ as follows:
\begin{equation}\label{impresp}
\begin{aligned}
\pi_{l}=h(\vec{r},t_0+(l-1)t_s)\big\rvert_{\|\vec{r}\|=d}, \qquad l\geq 1.
\end{aligned}
\end{equation}
According to the physical model, the number of counted molecules of type $\msM_i$ at the relay in the $k$-th super time slot, noted by $Y_{i,k}^\msR$, has a Poisson distribution with parameter $C_{i,k}v_\textrm{r}$, and the number of counted molecules of type $\msM_3$ at $\msT_i$, noted by $Y_{3,k}^{\msT_i}$ has a Poisson distribution with parameter $C_{3,k}v_\textrm{r}$.

%\vspace{-0.5em}
\subsection{The Proposed PNC Scheme}

Here, we propose a new PNC scheme based on chemical reactions in the medium, which makes the physical-layer XORing possible by exploiting the reaction among different molecule types and thus it does not need a logic XOR gate at the relay. In addition, by XORing at the physical-layer using reaction, the signal density reduces when both molecules arrive at the relay and thus the signal dependent noise at the relay is reduced. 

%we propose to implement the physical-layer XOR using the molecular reaction. Thus,
We first choose two molecule types $\msM_1$ and $\msM_2$, to be sent by the transceivers ($\msT_1$ and $\msT_2$, respectively), such that they can react with each other
% in the presence of a molecules type $\msE$
by a reversible reaction as follows:
\begin{equation}\label{ABreaction}
\msM_1+\msM_2 \underset{\kappa}{\overset {\gamma}{\rightleftharpoons}} \msM_{12},
\end{equation}
where $\gamma, \kappa \geq 0$ are the forward and reverse reaction rate constants of the molecules of type $\msM_1$ and $\msM_2$, respectively.
%molecules of type $\msE$ are released by the relay to assure that the molecules of type $\msM_1$ ans $\msM_2$ react with each other only around the relay.
The molecules of type $\msM_{12}$ cannot be detected by the receptors of the relay. The two communication phases in this scheme are similar to the SNC scheme with the difference that the XOR is performed in the medium instead of the relay and the relay implicitly decodes the physically made XOR of the messages and sends it to the transceivers in the second phase.
If both messages of the transmitters are $1$, both molecules $\msM_1$ and $\msM_2$ arrive at the relay and react with each other as in \eqref{ABreaction}. As a result, the concentrations of both molecules decrease in the environment and almost no molecule is measured by the receptors of the relay. When only $\msM_1$ or $\msM_2$ arrives at the relay, it is measured by its corresponding receptor at the relay. The stimulated receptor would release $\zeta_3$ molecules of type $\msM_3$ in the next time slot. Thus, to make a physical-layer XOR, it is enough to choose the number of released molecules appropriately. The block diagram of the system in the PNC scheme is the same as the SNC scheme (see Fig. 2) with the difference that the XOR is performed using reaction among molecules in the channel of the first communication phase (Phase 1 Channel). Further, instead of the XOR gate in the SNC, the decoded message of each receptor is encoded independently using OOK modulation at the encoder of the relay and the output signals add up naturally in the medium.

\textbf{Received concentration:} The physical model of the PNC scheme is similar to the SNC scheme, with the difference that in the PNC, \eqref{eqCiknoISI} is the concentration of molecules of type $\msM_i$ around the relay \emph{before reaction}, i.e., the concentrations of molecules of types $\msM_1$ and $\msM_2$ around the relay before reaction are $B_{1,k} \zeta_1 \pi_1$ and $B_{2,k} \zeta_2 \pi_1$, respectively. Similar to the SNC, we assume Poisson distribution for the number of counted molecules. In \cite{Schnoerr2016}, the authors show that assuming the Poisson distribution for the number of received molecules when there is reaction among molecules is an approximation which has excellent accuracy. This fact is also verified in \cite{Jamali2017} using particle-based simulation.
To find the received concentration of each molecule type at the relay, we obtain the concentration of molecules after reaction using the following reaction-diffusion equation:
\begin{equation}\label{eqreact}
\begin{aligned}
\frac{\partial c_1(\vec{r},t)}{\partial t} = D_1 \nabla^2 c_1(\vec{r},t)
-\gamma c_1(\vec{r},t) c_2(\vec{r},t)+\kappa c_{12}(\vec{r},t)+g_1(\vec{r},t),\\
\frac{\partial c_2(\vec{r},t)}{\partial t} = D_2 \nabla^2 c_2(\vec{r},t)
-\gamma c_1(\vec{r},t) c_2(\vec{r},t)+\kappa c_{12}(\vec{r},t)+g_2(\vec{r},t),
%\\ \frac{\partial c_{12}(\vec{r},t)}{\partial t} = D_{12} \nabla^2 c_{12}(\vec{r},t)
%+\gamma c_1(\vec{r},t) c_2(\vec{r},t)-\kappa c_{12}+g_2(\vec{r},t),
\end{aligned}
\end{equation}
where $c_{12}(\vec{r},t)$ is the concentration of molecule type $\msM_{12}$. Since $D_1=D_2=D$, subtracting the equations in \eqref{eqreact}, the nonlinear term cancels out and we obtain
\begin{equation}\label{eqreactdiff}
\begin{aligned}
\frac{\partial \rho(\vec{r},t)}{\partial t} = D \nabla^2 \rho(\vec{r},t)+g_1(\vec{r},t)-g_2(\vec{r},t),
\end{aligned}
\end{equation}
where $\rho(\vec{r},t)=c_1(\vec{r},t)-c_2(\vec{r},t)$. Hence, the difference of molecule densities can be obtained using superposition property for LTI systems as $\rho(\vec{r},t)=g_1(\vec{r},t)*h(\vec{r},t)-g_2(\vec{r},t)*h(\vec{r},t)$, where $h(\vec{r},t)$ is given in \eqref{impulseresp} (this is a key argument of \cite{IWCIT2016type}). However, there is no general closed-form expression for the concentration of each molecule type because of the nonlinear term. To address this difficulty and obtain closed-form expressions for our analysis, from now on, we assume \emph{perfect reaction} among molecules of types $\msM_1$ and $\msM_2$ (i.e., forward reaction rate constant $\gamma$ goes to $\infty$ and reverse reaction rate constant $\kappa$ is zero).\footnote{The PNC scheme with imperfect reaction is considered in Section \ref{Simulation} and the concentration of molecules is obtained numerically using the finite difference method.} In this case, when the two molecule types meet in the medium, molecule type with lower concentration is completely canceled out, and a residual part of the one with higher concentration remains, \emph{i.e.}, both molecule types do not exist simultaneously in the same location in the medium ($c_1(\vec{r},t)c_2(\vec{r},t)=0$). With this assumption, the concentration of each molecule type can be obtained as follows: if $\rho(\vec{r},t)\geq 0$, $c_2(\vec{r},t)=0$ and $c_1(\vec{r},t)=\rho(\vec{r},t)$; otherwise, $c_1(\vec{r},t)=0$ and $c_2(\vec{r},t)=-\rho(\vec{r},t)$. In other words, $c_1(\vec{r},t)=\max\{0,\rho(\vec{r},t)\}$ and $c_2(\vec{r},t)=\max\{0,-\rho(\vec{r},t)\}$.

When we have no ISI, the concentration of molecules of type $\msM_1$ and $\msM_2$ measured by the receptors of their type are $C_{1,k}=\max\{B_{1,k} \zeta_1 \pi_1-B_{2,k} \zeta_2 \pi_1,0\}$ and $C_{2,k}=\max\{B_{2,k} \zeta_2 \pi_1- B_{1,k} \zeta_1 \pi_1,0\}$, respectively. Since we assumed the same channel coefficients for the two transceivers, if we choose $\zeta_1=\zeta_2=\zeta$, an almost equal concentration of molecules of both types arrives at the relay (when both transceivers send the information bit $1$).\footnote{If the channel coefficients are not equal, each transceiver $\msT_i$, knowing its own channel coefficient $\pi_1$, chooses $\zeta_i$ such that a constant concentration of molecules arrive at the relay when the message of the transceiver is 1). }
This makes almost all molecules react with each other and thus realizes a physical-layer XOR.
For a fair comparison of the two schemes, from now on, we assume $\zeta_1=\zeta_2=\zeta$, for both schemes.

The error performances of the two schemes without ISI are investigated in Section \ref{error}. It is shown analytically and later by simulation that the proposed PNC scheme outperforms the SNC scheme.

\textbf{Remark on notation:} While the notation is kept as simple as possible, the messages sent and decoded by the transceivers and the relay in each communication phase must be defined as well as the error events for each phase. Furthermore, since we have two receptor types at the relay, the decoded messages of each receptor type and their corresponding error events must be defined. Table \ref{table_notation} summarizes the mostly used notations in this paper.

\begin{table}
\centering
\caption{Used Notations}
\begin{tabular}{|p{2.9cm}|p{12.7cm}|}
\hline
$B_{i,k}$ & The message of the transceiver $\msT_i$ in the $k$-th super time slot
\\
$B_{\msR,k}$ &The sent message of the relay in the $k$-th super time slot
\\
$B_{\msR_i,k}$ & A part of the message $B_{\msR,k}$, to be decoded by the $i$-th receptor at the relay in the PNC in the $k$-th\\
$=B_{i,k} \cdot (B_{1,k} \oplus B_{2,k})$ & super time slot
\\
$\hat{B}_{\msR_i,k}$ & The decoded message by the $i$-th receptor at the relay in the PNC in the $k$-th super time slot
\\
$\hat{B}_{i,k}^{\msR}$ & The decoded message by the $i$-th receptor at the relay in the SNC
in the $k$-th super time slot
\\
$\hat{B}_{\msR,k}^{\msT_i}$ & The message of the relay, decoded at the transceiver $\msT_i$
in the $k$-th super time slot
\\
$\hat{B}_{i,k}^{\msT_j}$ & The message of the transceiver $\msT_i$, decoded by the transceiver $\msT_i$ in the $k$-th super time slot\\
$X_{i,k}$ & The number of released molecules of type $\msM_i$ in the $k$-th super time slot\\
$C_{i,k}$ & The concentration of molecules of type $\msM_i$ around its receptors in the $k$-th super time slot\\
$I_{i,k}$ & The concentration of remained molecules of type $\msM_i$ from the previous super time slots around its receptors in the $k$-th super time slot\\
$E_{i,k}$ & The error event at transceiver $\msT_i$ in the $k$-th super time slot ($\hat{B}_{\bar{i},k}^{\msT_i} \neq B_{\bar{i},k}$)\\
$E_{\msR,k}$ & The error event of the first communication phase in the $k$-th super time slot ($B_{\msR,k} \neq B_{1,k} \oplus B_{2,k}$)\\
$E_{\msR_i,k}$ & The error event of the $i$-th receptor at the relay in the $k$-th super time slot (in PNC: $\hat{B}_{\msR_i,k} \neq B_{\msR_i,k}$, in SNC: $\hat{B}_{i,k}^{\msR} \neq B_{i,k}$)\\
$E_k^{\msT_i}$ & The error event of the second communication phase in the $k$-th super time slot ($\hat{B}_{\msR,k}^{\msT_i} \neq B_{\msR,k}$)\\
\hline
\end{tabular}
\label{table_notation}
\vspace{-1.5em}
\end{table}

%\vspace{-0.5em}
\section{Chemical Reaction for Dual Purpose of Transmission}\label{sec:ISImitigate}
In this section, to illustrate the idea of the dual purpose of transmission (as mentioned in the introduction), we consider the ISI case and using the reaction characteristic of the PNC scheme, we propose an ISI mitigating technique for the first communication phase of the PNC scheme. To have a fair comparison between the two schemes, we apply the existing ISI mitigating techniques to the SNC scheme. In our schemes, we assume that the transceivers know the channel coefficients of both transceivers to the relay, i.e., the distances and diffusion coefficients.

In Subsection~\ref{sec:SNC_ISI}, ISI mitigating technique for SNC scheme is described, and in Subsection~\ref{sec:PNC_ISI} the proposed ISI mitigating technique for PNC is presented.

%\vspace{-0.5em}
\subsection{The SNC Scheme}\label{sec:SNC_ISI}

In the SNC scheme, we use the existing ISI mitigating techniques (as mentioned, SNC in the presence of ISI has not been studied before). To mitigate ISI in a communication link, two approaches are possible: adapting transmission rate at the transmitter \cite{movahednasab2015}, and adapting threshold at the receiver \cite{mosayebi2014}.
In both communication phases, to reduce the complexity of the relay, we leave all complexity to the transceivers. 

In communication phase 1, we use an adaptive rate\footnote{From now on by "adaptive rate", we mean "adaptive transmission rate"} at the transceivers along with a fixed threshold at the relay. This means that we extend the method of \cite{movahednasab2015} to the SNC scheme, i.e., each transceiver adapts its transmission rate to mitigate its own ISI.
Therefore, when the message of the transceiver is $0$, it stays silent; otherwise, according to its transmission in the previous super time slot, it adapts its rate such that the concentration of molecules around the relay is a constant value. More specifically, in the $k$-th super time slot, if $B_{i,k}=0$, the transceiver $\msT_i$ stays silent and if $B_{i,k}=1$, the transceiver transmits an adaptive number of molecules such that a constant concentration of molecules, $c_\textrm{r}$, arrives at the relay. Hence, each $\msT_i$ to send its message $B_{i,k} \in \{0,1\}$ in the $k$-th super time slot transmits
\begin{equation}\label{eqXikSNCadap}
\begin{aligned}
X_{i,k}=B_{i,k}(\frac{c_\textrm{r}}{\pi_1}-L_{i,k}),
\end{aligned}
\end{equation}
molecules such that
\begin{equation}\label{eq_LTikSNC}
\begin{aligned}
L_{i,k}=\frac{I_{i,k}}{\pi_1}.
\end{aligned}
\end{equation}
where $I_{i,k}$ denotes the ISI term, which is the concentration of molecules of type $\msM_i$ around the relay remained from the previous super time slots. Note that in this scheme, the ISI is not mitigated when the message of the transceiver is 0.
In the following, we first describe the ISI model and obtain the concentration of received molecules at the relay in the presence of ISI. Then, using the received concentration, we obtain the adaptive rates in Lemma \ref{lemma1}.

\textbf{Received concentration:} We model the ISI in the channel by a $q$-slot memory\cite{Arjmandi2014}, i.e., $\pi_l=0$, for $l>q+1$, where $\pi_l$ is defined in \eqref{impresp}. In addition, since in our transmission protocol, the molecules of types $\msM_1$ and $\msM_2$ are released in the odd time slots and the molecules of type $\msM_3$ are released in the even time slots, the relay counts the number of received molecules in the odd time slots and the transceivers count the number of received molecules in the even time slots. Hence, the number of received molecules is counted once in every two time slots at the receivers of the relay and the transceivers. Therefore, the ISI in the $l$-th time slot is caused by the concentration of remained molecules from $(l-2)$-th, $(l-4)$-th,..., $(l-2\lfloor \frac{q}{2}\rfloor)$-th time slots. For example, for $q=1$, when the relay counts the number of received molecules in the $l$-th time slot, no ISI is observed, because the transceivers have not released any molecules in the $(l-1)$-th time slot. Hence, the concentration of molecules of type $\msM_i$ around the relay in the $k$-th super time slot is given as
\begin{equation}\label{eqCik}
\begin{aligned}
C_{i,k}=\sum_{l=0}^{\lfloor \frac{q}{2}\rfloor} \pi_{2 l+1} X_{i,k-l}=X_{i,k} \pi_1+I_{i,k},\qquad i \in \{1,2\},
\end{aligned}
\end{equation}
%where $I_{i,k}$ denotes the ISI term, which is the concentration of molecules of type $\msM_i$ around the relay remained from the previous super time slots. The concentration of molecules around the transceivers can be obtained similarly.

\begin{lemma} \label{lemma1}The adaptive rate of the transceiver $\msT_i$ in the SNC scheme to mitigate ISI of the transceiver-relay channel with $q$-slot memory, can be obtained as follows:
\begin{align}\label{XikSNCbargashti}
X_{i,k}=B_{i,k}\big(\frac{c_\textrm{r}}{\pi_1}-\sum_{l=1}^{\lfloor \frac{q}{2} \rfloor}\eta_{2l+1} X_{i,k-l}\big),\qquad i \in\{1,2\},
\end{align}
where $\eta_l$ is the normalized channel gain as follows:
\begin{align}\label{eqnui}
\eta_l=\frac{\pi_l}{\pi_1}, \qquad l>1.
\end{align}
\end{lemma}
\begin{proof} The proof is straightforward by substituting \eqref{eq_LTikSNC} in \eqref{eqXikSNCadap} and using the fact that for $q$-slot memory in the channel, $I_{i,k}=\sum_{l=1}^{\lfloor \frac{q}{2}\rfloor} \pi_{2 l+1}X_{i,k-l}$ from \eqref{eqCik}.
\end{proof}
\begin{remark}
According to \eqref{XikSNCbargashti}, each transceiver $\msT_i$ needs to save the number of its released molecules in previous $\lfloor \frac{q}{2} \rfloor$ super time slots, i.e., $X_{i,k-1},...X_{i,k-\lfloor \frac{q}{2} \rfloor}$, to determine $X_{i,k}$. Note that the number of released molecules from $\msT_i$ in each super time slot has a maximum value which can be obtained from \eqref{XikSNCbargashti} when $X_{i,k-1}=...=X_{i,k-\lfloor \frac{q}{2} \rfloor}=0$ and $B_{i,k}=1$ as $X_{i,\textrm{max}}=\frac{c_\textrm{r}}{\pi_1}$. Hence, a finite memory is needed to save $X_{i,k-l}$. Further, we assume $\eta_l<1$, for $l>1$.\footnote{In diffusion-based systems with channel memory, the sampling time $t_0$ is chosen such that $h(d,t)$ takes its maximum at $t=t_0$, and thus $\pi_{l}<\pi_1$, for $l>1$. Hence, to make the channel coefficients to be reducing, we choose $t_0=\frac{d^2}{6D}$ \cite{Llatser2013}.} Note that when $B_{i,k}=1$, from \eqref{XikSNCbargashti} we have, $X_{i,k}\geq \frac{c_\textrm{r}}{\pi_{1}}-\sum_{l=1}^{\lfloor \frac{q}{2} \rfloor}\eta_{2l+1}X_{i,\textrm{max}}=\frac{c_\textrm{r}}{\pi_{1}}(1-\sum_{l=1}^{\lfloor \frac{q}{2} \rfloor}\eta_{2l+1})$. Hence, if the channel coefficients are such that $\sum_{l_2=1}^{\lfloor\frac{q}{2}\rfloor}\eta_{2l_1+1} < 1$,\footnote{This condition can be assured at the cost of increasing $t_s$ and decreasing the sampling rate accordingly. This decreases the values of $\eta_l$.} it is assured that $X_{i,k}$ is always non-negative.
\end{remark}

In communication phase 2, the second ISI mitigating approach, \emph{i.e.}, using adaptive thresholds at the transceivers along with fixed rate at the relay, is taken \cite{mosayebi2014}. The adaptive thresholds of phase 2 are derived in Section~\ref{errorISI}.

%\vspace{-0.5em}
\subsection{The Proposed PNC Scheme}\label{sec:PNC_ISI}
In the PNC scheme, the XOR is realized in the medium using the molecular reaction in the first communication phase. In the presence of ISI, there are remaining molecules from the previous transmissions. Using the idea of the dual purpose of transmission, we use reaction to mitigate ISI in the first communication phase by releasing extra molecules from each transceiver to react with the remaining molecules of the other transceiver from the previous transmissions. Note that each transceiver has access to the decoded version of the transmitted bits of the other transceiver in the previous super time slots. Thus, knowing its own channel coefficients and the channel coefficients of the other transceiver, each transceiver can estimate the concentration of the remaining molecules of the other transceiver from the previous transmissions. Therefore, it can choose its transmission rate such that along with transmitting its own message, the concentration of the remaining molecules of the other transceiver is also canceled out. As an example, assume the case of one super time slot memory for the transceiver-relay channel. Also assume the messages of the transceivers $\msT_1$ and $\msT_2$ to be $1$ and $0$, respectively, in the current super time slot. Because of the one super time slot memory in the channel, there may be concentrations of the remaining molecules of types $\msM_1$ and $\msM_2$ around the relay from the previous super time slot. The transceiver $\msT_1$ releases a constant number of molecules to send its information bit $1$ and some extra molecules to cancel out the remaining molecules of the transceiver $\msT_2$ from the previous super time slot. Since the message of $\msT_2$ is $0$, it does not release any molecules to send its message, but releases some molecules to cancel out the remaining molecules of $\msT_1$ around the relay from the previous super time slot. Hence, in this scheme, the transceivers may release some molecules even if their message is $0$. This scheme is based on using an adaptive rate at the transceivers along with a fixed threshold at the relay, where similar to SNC, the relay is assumed to be simple while all complexity is left to the transceivers. 
%The transceivers do not know the very exact number of the released molecules of the other transceiver in the previous super time slots, but can estimate it. We show that, for a unit super time slot memory, each transceiver can use the decoded message of the other transceiver in the $(k-1)$-th super time slot and the number of its own released molecules in the $(k-2)$-th super time slot to estimate the number of the released molecules of the other transceiver in the previous super time slot. %The ISI-mitigated PNC scheme is explained in detail. 
More specifically, each transceiver $\msT_i$ releases extra molecules, denoted by $U_{i,k}$, in each super time slot to react with and cancel out the remained molecules of the other transceiver from the previous super time slots, i.e., for $i\in \{1,2\}$,
\begin{align}\label{eqXikPNCadap}
X_{i,k}= B_{i,k}\frac{c_\textrm{r}}{\pi_1}+U_{i,k},
\end{align}
in which
\begin{align}\label{eq_LTikPNC}
U_{i,k}=\frac{\tilde{I}_{\bar{i},k}}{\pi_1},
\end{align}
where $\tilde{I}_{\bar{i},k}$ is the estimated value of the remained molecules of the transceiver $\msT_{\bar{i}}$ around the relay in the $k$-th super time slot before reaction, which is calculated by the transceiver $\msT_i$ using its previously decoded messages. Note that, despite the SNC scheme, in this scheme, the ISI of the channel is mitigated for both messages 0 and 1. The ISI model in this scheme is similar to the SNC scheme, with the difference that in the PNC, \eqref{eqCik} is the concentration of molecules of type $\msM_i$ around the relay before reaction. The adaptive rates is obtained in Lemma \ref{lemma2}.

%Similar to the SNC scheme, we first assume $\lfloor\frac{q}{2}\rfloor=1, i =1,2$ and then extend it to higher channel memories. The adaptive rates for one super time slot memory is obtained in Lemma \ref{lemma2}.
\begin{lemma}\label{lemma2} The adaptive rate of the transceiver $\msT_i$ in the PNC scheme to mitigate ISI of the transceiver-relay channels with $q$-slot memory, can be obtained as follows:
\begin{align}\label{XikPNCbargashti}
X_{i,k}=\frac{c_\textrm{r}}{\pi_1}(B_{i,k}+\sum_{l=1}^{\lfloor\frac{q}{2}\rfloor}\eta_{2l+1} \hat{B}_{\bar{i},k-l}^{\msT_i})+\sum_{l_1=1}^{\lfloor\frac{q}{2}\rfloor}\sum_{l_2=1}^{\lfloor\frac{q}{2}\rfloor}\eta_{2l_1+1} \eta_{2l_2+1} X_{i,k-l_1-l_2},\qquad i \in \{1,2\},
\end{align}
\end{lemma}
\begin{proof}
By substituting \eqref{eq_LTikPNC} in \eqref{eqXikPNCadap} and using $\tilde{I}_{\bar{i},k}=\sum_{l=1}^{\lfloor \frac{q}{2}\rfloor} \pi_{2 l+1}\tilde{X}_{\bar{i}, k-l}$ (where $\tilde{X}_{\bar{i}, k-l}$ is the approximated value of the number of released molecules from $T_{\bar{i}}$ in the $(k-l)$-th super time slot, calculated by $\msT_i$ using its decoded messages), we have:
\begin{align}\label{XikPNCbargashti_1}
X_{i,k}= B_{i,k}\frac{c_\textrm{r}}{\pi_1}+\sum_{l_1=1}^{\lfloor \frac{q}{2}\rfloor} \eta_{2 l_1+1}\tilde{X}_{\bar{i}, k-l_1},
\end{align}
for $ i \in \{1,2\}$. We can write a similar equation for $\tilde{X}_{\bar{i},k-l_1}$ as follows:
\begin{align}\label{XikPNCbargashti_2}
\tilde{X}_{\bar{i},k-l_1}=\hat{B}_{\bar{i},k-l_1}^{\msT_i}\frac{c_\textrm{r}}{\pi_1}+\sum_{l_2=1}^{\lfloor \frac{q}{2}\rfloor} \pi_{2 l_2+1}X_{i, k-l_1-l_2}.
\end{align}
Now, by substituting \eqref{XikPNCbargashti_2} in \eqref{XikPNCbargashti_1}, we obtain \eqref{XikPNCbargashti}.
\end{proof}
\begin{remark}
According to \eqref{XikPNCbargashti},
each transceiver $\msT_i$ needs to save its decoded messages in previous $\lfloor \frac{q}{2} \rfloor$ super time slots
%, i.e., $\{\hat{B}_{\bar{i},k-1}^{\msT_i},...,\hat{B}_{\bar{i}}^{\msT_i,k-\lfloor \frac{q}{2} \rfloor}\}$
and the number of its released molecules in previous
$2\lfloor\frac{q}{2} \rfloor$ super time slots. If the channel coefficients are such that $\sum_{l_2=1}^{\lfloor\frac{q}{2}\rfloor}\eta_{2l_1+1} < 1$, it can be easily shown by induction that the number of released molecules from $\msT_i$ in each super time slot is always bounded, i.e., $X_{i,k} < \frac{c_\textrm{r}}{\pi_1} \cdot \frac{1}{1-\sum_{l=1}^{\lfloor\frac{q}{2}\rfloor}\eta_{2l+1}}$.
This guarantees the stability of the scheme.
\end{remark}

\begin{remark} In Section \ref{Simulation}, for a fair comparison of the SNC and PNC schemes, we choose $c_\textrm{r}^\textrm{SNC}$ (the $c_\textrm{r}$ used in \eqref{eqXikSNCadap}) and $c_\textrm{r}^\textrm{PNC}$ (the $c_\textrm{r}$ used in \eqref{eqXikPNCadap}), such that
%the average numbers of the released molecules from the two transceivers are equal in the two schemes, i.e.,
%\begin{align}\label{eqfair}
$\frac{1}{2}\sum_{i =1}^{2}X_{i,\textrm{avg}}^{\textrm{SNC}}=\frac{1}{2}\sum_{i=1}^{2} X_{i,\textrm{avg}}^{\textrm{PNC}},$
%\end{align}
where $X_{i,\textrm{avg}}^{\textrm{PNC}}$ and $X_{i,\textrm{avg}}^{\textrm{SNC}}$ are the average number of the released molecules from the transceiver $\msT_i$ in the PNC and SNC schemes, respectively. The average values can be obtained from \eqref{XikSNCbargashti} and \eqref{XikPNCbargashti} by substituting $X_{i,k}$ and $X_{i,k-1}$ with their average values ($X_{i,\textrm{avg}}^\textrm{PNC}$ or $X_{i,\textrm{avg}}^\textrm{SNC}$), and $B_{i,k}$ and $B_{\bar{i},k-1}$ with their average values, $\frac{1}{2}$, as follows:
\begin{align}
&X_{i,\textrm{avg}}^\textrm{SNC}=\frac{c_\textrm{r}^\textrm{SNC}}{\pi_1} \cdot \frac{1}{2+\sum_{l=1}^{\lfloor \frac{q}{2} \rfloor}\eta_{2l+1}}, \quad X_{i,\textrm{avg}}^\textrm{PNC}=\frac{c_\textrm{r}^\textrm{PNC}}{\pi_1} \cdot \frac{1}{2(1-\sum_{l=1}^{\lfloor\frac{q}{2}\rfloor}\eta_{2l+1})}, \qquad i \in \{1,2\}.
\end{align}
Hence,
\begin{align}
&\frac{c_\textrm{r}^\textrm{SNC}}{c_\textrm{r}^\textrm{PNC}}=\frac{2+\sum_{l=1}^{\lfloor\frac{q}{2}\rfloor}\eta_{2l+1}}{2(1-\sum_{l=1}^{\lfloor\frac{q}{2}\rfloor}\eta_{2l+1})}.
\end{align}
\end{remark}

%The error performance of the two schemes in the presence of ISI using the proposed ISI mitigating techniques are investigated in Section \ref{errorISI} for unit memory. It is seen by simulation that the error performance of the PNC using the proposed ISI mitigating technique is much better than the SNC using the existing ISI mitigating technique.
In communication phase 2, similar to SNC, the approach of adapting thresholds at the receivers is used with the thresholds derived in Section~\ref{errorISI}.

\section{Error Performance Analysis with No ISI}\label{error}
In this section, we derive the probabilities of error at the transceivers $\msT_1$ and $\msT_2$, noted by $p_{\textrm{e},1}$ and $p_{\textrm{e},2}$, respectively. Throughout this paper, we consider the average bit error probability (Avg-BEP) as follows:
\begin{equation}\label{AvgBEP}
\textrm{Avg-BEP}=\frac{1}{2}(p_{\textrm{e},1}+p_{\textrm{e},2}).
\end{equation}
First, we investigate the error probabilities of the SNC scheme. Then, using a similar approach, we derive the error probabilities of the proposed PNC scheme. Since the error probability without ISI in the current super time slot does not depend on the error probabilities of the previous super time slots and is the same for all super time slots, we drop the index $k$ of the bits and error events in this section.

\subsection{The SNC scheme}
In the SNC scheme, as mentioned before, the $i$-th receptor at the relay decodes $B_i$ (the message of the transceiver $\msT_i$) as $\hat{B}_i^{\msR}$. The relay XORs the decoded messages and sends the message $B_{\msR}=\hat{B}_1^{\msR} \oplus \hat{B}_2^{\msR}$ to the transceivers using $X_3^\msR=B_{\msR}\zeta_3$ molecules of type $\msM_3$. Each transceiver $\msT_i$ decodes the message of the relay as $\hat{B}_{\msR}^{\msT_i}$ and, finds the message sent by the other transceiver as $\hat{B}_{\bar{i}}^{\msT_i}=B_i \oplus \hat{B}_{\msR}^{\msT_i}, i \in\{1,2\}$.

Define $E_i$ as the error event at the transceiver $\msT_i$, i.e., {$\hat{B}_{\bar{i}}^{\msT_i} \neq B_{\bar{i}}$}. The probability of the event $E_i$ is shown by $\mP(E_i)=p_{\textrm{e},i}$. $E_i$ consists of two error events corresponding to two communication phases:

i. $E_{\msR}$: $B_{1} \oplus B_{2}$ is decoded with error at the relay ($B_{\msR} \neq B_{1} \oplus B_{2} $).

ii. $E^{\msT_i}$: The $i$-th transceiver decodes the message of the relay with error ($\hat{B}_{\msR}^{\msT_i} \neq B_{\msR}$).\\
The probabilities of the first and second events are shown by $\mP(E_{\msR})$ and $\mP(E^{\msT_i})$, respectively. We show the conditioned event $\{B=b\}$ with $\{b\}$ for brevity, when it is clear from the context. The total error probability at the transceiver $T_i$ in terms of $\mP(E_{\msR})$ and $\mP(E^{\msT_i})$ is obtained in Lemma \ref{lemmaPei}.

\begin{lemma}\label{lemmaPei} The total error probability at the transceiver $\msT_i$ in the SNC scheme can be obtained as
\begin{align}\label{eq_Peim1m2}
p_{\textrm{e},i}=
%\mP(E_i)=\mP(E_{\msR})(1-\mP(E^{\msT_i}))+(1-\mP(E_{\msR}))\mP(E^{\msT_i})
\frac{1}{2}\sum_{b_R \in \{0,1\}}\mP(E^{\msT_i}|B_{\msR}=b_\msR)+\frac{1}{4}\big[1-\sum_{b_R \in \{0,1\}}\mP(E^{\msT_i}|B_{\msR}=b_\msR)\big]\sum_{b_1,b_2 \in \{0,1\}} \mP(E_{\msR}|b_1,b_2).
\end{align}
\end{lemma}
\begin{proof} The total error probability at $\msT_i$ conditioned to $B_1=b_1$ and $B_2=b_2$ can be obtained as
\begin{align}\label{eq_Pei}
\nonumber
\mP(E_i|&B_1=b_1,B_2=b_2)=\mP(\hat{B}_{\bar{i}}^{\msT_i}\neq B_{\bar{i}}|b_1,b_2)\\\nonumber
=&\mP(\hat{B}_{\msR}^{\msT_i}=B_{\msR}, B_{\msR} \neq b_1 \oplus b_2|b_1,b_2)+\mP(\hat{B}_{\msR}^{\msT_i} \neq B_{\msR}, B_{\msR}=b_1 \oplus b_2|b_1,b_2)\\
=&\mP(E_{\msR}|b_1,b_2)\big(1-\mP(E^{\msT_i}|B_{\msR}=\overline{b_1 \oplus b_2})\big)+\big(1-\mP(E_{\msR}|b_1,b_2)\big)\mP(E^{\msT_i}|B_{\msR}=b_1 \oplus b_2),
\end{align}
for $i \in \{1,2\}$. By taking average over $B_1$ and $B_2$, the total probability of error at the transceiver $\msT_i$ can be easily obtained for $i \in \{1,2\}$ as \eqref{eq_Peim1m2}.
%\begin{align}\nonumber
%\mP(E_i)&=\frac{1}{4}\sum_{b_1,b_2 \in \{0,1\}} \mP(E_i|B_1=b_1,B_2=b_2)\\\nonumber
%&=\mP(E_{\msR})+\mP(E^{\msT_i})-\frac{1}{4}\sum_{b_1,b_2 \in \{0,1\}}\mP(E_{\msR}|b_1,b_2)(\mP(E^{\msT_i}|B_{\msR}=\overline{b_1 \oplus b_2})+\mP(E^{\msT_i}|B_{\msR}=b_1 \oplus b_2))\\
%&=\mP(E_{\msR})+\mP(E^{\msT_i})-2\mP(E_{\msR})\mP(E^{\msT_i}).
%\end{align}
\end{proof}
In the following, we compute the error probabilities of the two phases, i.e., $\mP(E_{\msR}|b_1,b_2)$ and $\mP(E^{\msT_i}|B_{\msR})$.

\textbf{Phase 1:} When $\msT_i$ sends $B_i=b_i$, the average concentration molecules of type $\msM_i$ at the relay is $C_i=b_i\zeta \pi_1$. Hence, due to the transparent receiver, the number of counted molecules of type $\msM_i$ at the relay, $Y_i^\msR$, is Poisson$(b_i\zeta \pi_1v_\textrm{r})$. The relay uses a threshold $\tau_i^{\msR}$ to decode $B_i$: if $Y_i\leq \tau_i^{\msR}$, then $B_i$ is decoded as $\hat{B}_i^{\msR}=0$; otherwise, it is decoded as $\hat{B}_i^{\msR}=1$.

\begin{lemma}\label{lemmathreshold}
For a transparent receiver with no environment noise, the optimum threshold to decode a message $B \in \{0,1\}$ with $\mP\{B=0\} \geq \mP\{B=1\}$,\footnote{While assumption of uniform distribution on message bits is common in communication systems, we need this general condition to obtain the optimum thresholds at the receivers (both at the relay and the transceivers). Note that our assumption allows for the uniform distribution on message bits.} which is sent using OOK modulation, is obtained using the maximum-a-posteriori (MAP) decision rule as $\tau=0$.
\end{lemma}
\begin{proof}For a transparent receiver with no environment noise, if $Y$ is the number of counted molecules at the receiver, we have
\begin{align}\label{delta}
\mP{\{Y=y|B=0\}}=\delta[y].
\end{align}
Using the MAP decision rule to decode $B$, we have
\begin{align}\label{threshold1}
\mP\{B=1\}\mP{(y|B=1)}\overset{1}{\underset{0}\gtrless}\mP{\{B=0\}}\mP{(y| B=0)}=\mP\{B=0\}\delta[y],
\end{align}
which results in $\tau=0$, since $\mP\{B=0\}\geq\mP\{B=1\}$.
\end{proof}

Using Lemma \ref{lemmathreshold} and because of the uniform distribution of the message bits of the transceivers, the optimum thresholds at the relay are $\tau_i^\msR=0$, $i \in \{1,2\}$. Using these thresholds at the relay, the error probability of the first phase is obtained in Lemma \ref{lemmaPeR}.

\begin{lemma}\label{lemmaPeR} The error probability of the first communication phase, $\mP(E_{\msR}|b_1,b_2)$, in the SNC scheme using the optimum threshold can be obtained as
\begin{align}\label{eq_PeRSNC}
\mP(E_{\msR}|B_1=0,B_2=0)&=0,\\\nonumber
\mP(E_{\msR}|B_1=1,B_2=0)&=\mP(E_{\msR}|B_1=0,B_2=1)=\exp(-\zeta\pi_1v_\textrm{r}),\\\nonumber
\mP(E_{\msR}|B_1=1,B_2=1)&=2\exp(-\zeta\pi_1v_\textrm{r})-2\exp(-2\zeta\pi_1v_\textrm{r}).
\end{align}
\end{lemma}
\begin{proof} We define $E_i^{\msR}=\{\hat{B}_i^{\msR} \neq B_i\}$ to denote the event where $B_i$ is decoded with error at the relay. Note that $B_i=0$ is decoded without error at the relay, due to the noiseless assumption. Hence, for $i\in\{1,2\}$
\begin{equation}\label{eq_PeT1type}
\begin{aligned}
\mP(E_i^{\msR}|B_i=0)&=\mP\{Y_i^{\msR}> \tau_i^\msR|B_i=1\}=0,\\
\mP(E_i^{\msR}|B_i=1)&=\mP\{Y_i^{\msR}\leq \tau_i^\msR|B_i=1\}=\exp(-\zeta\pi_1v_\textrm{r}).
\end{aligned}
\end{equation}
Due to XORing at the relay, the event $E_{\msR}$ is equivalent to the event that one of the messages $B_1$ or $B_2$ is decoded with error at the relay. Hence,
\begin{align}\label{eq_PeR}
\mP(E_{\msR}|B_1=b_1,B_2=b_2)=\mP(E_1^{\msR}|b_1)(1-\mP(E_2^{\msR}|b_2))+(1-\mP(E_1^{\msR}|b_1))\mP(E_2^{\msR}|b_2).
\end{align}
By substituting $\mP(E_1^{\msR}|b_1)$ and $\mP(E_2^{\msR}|b_2)$ from \eqref{eq_PeT1type} in \eqref{eq_PeR}
%we obtain:
%\begin{align}\label{eq_PeRSNC_cond}
%\mP(E_{\msR}|B_1=0,B_2=0)&=0,\\\nonumber
%\mP(E_{\msR}|B_1=1,B_2=0)&=\mP(E_{\msR}|B_1=0,B_2=1)=\exp(-\zeta\pi_1v_\textrm{r}),\\\nonumber
%\mP(E_{\msR}|B_1=1,B_2=1)&=2\exp(-\zeta\pi_1v_\textrm{r})-2\exp(-2\zeta\pi_1v_\textrm{r}).
%\end{align}
%Finally, by taking average over $B_1$ and $B_2$,
we obtain \eqref{eq_PeRSNC}.
\end{proof}

\textbf{Phase 2:} The conditional distribution of the number of counted molecules of type $\msM_3$ at the transceiver $\msT_i$, i.e., $Y_3^{\msT_i}$, given $B_{\msR}=b_\msR$, is Poisson$\big(b_\msR\zeta_3 \pi_1v_\textrm{r}\big)$. To decode $B_{\msR}$, each transceiver $\msT_i$ uses a threshold, $\tau^{\msT_i}$. The optimum value of $\tau^{\msT_i}$ can be obtained according to Lemma \ref{lemmathreshold} as $\tau^{\msT_i}=0$, $i \in\{1,2\}$.\footnote{Note that $\mP\{B_{\msR}=0\}=\frac{1}{4}[\mP\{E_{\msR}^c|B_1=0,B_2=0\}+\mP\{E_{\msR}^c|B_1=1,B_2=1\}+\mP\{E_{\msR}|B_1=0,B_2=1\}+\mP\{E_{\msR}|B_1=1,B_2=0\}]=\frac{1}{2}(1+\exp(-2\zeta\pi v_\textrm{r}))$, which is greater than $\mP\{B_{\msR}=1\}=1-\mP\{B_{\msR}=0\}=\frac{1}{2}(1-\exp(-2\zeta\pi v_\textrm{r}))$ and thus the condition in Lemma \ref{lemmathreshold} is satisfied.} The error probability of the second phase is obtained in Lemma \ref{lemmaPeTi}.
\begin{lemma}\label{lemmaPeTi}
The error probability of the second communication phase at $\msT_i$, $\mP(E^{\msT_i}|b_\msR)$, in the SNC scheme using the optimum threshold can be obtained as
\begin{align}\label{eq_PeRihat}
&\mP(E^{\msT_i}|B_{\msR}=0)=0, \quad \mP(E^{\msT_i}|B_{\msR}=1)=\exp(-\zeta_3\pi_1v_\textrm{r}).
\end{align}
\end{lemma}
\begin{proof}
The proof is straightforward according to the fact that $\mP(E^{\msT_i}|B_{\msR}=0)=\mP\{Y_3^{\msT_i}>\tau^\msT_i|B_{\msR}=0)$ and $\mP(E^{\msT_i}|B_{\msR}=1\}=\mP\{Y_3^{\msT_i}\leq \tau^\msT_i|B_{\msR}=1\}$.
\end{proof}

Now, by substituting the error probabilities of the two communication phases (from $\eqref{eq_PeRSNC}$ and \eqref{eq_PeRihat}) in \eqref{eq_Peim1m2}, we obtain $p_{\textrm{e},i}$, $i \in \{1,2\}$, as
\begin{align}\label{Peitype1}
p_{\textrm{e},i}=\frac{1}{2}\exp(-\zeta_3\pi_1v_\textrm{r})+\frac{1}{2}\left[1-\exp(-\zeta_3\pi_1v_\textrm{r})\right]%\\\nonumber&\quad\times
 \left[2\exp(-\zeta\pi_1v_\textrm{r}) -\exp(-2\zeta\pi_1v_\textrm{r}) \right].
\end{align}

\subsection{The proposed PNC scheme}\label{error:PNC}
In the PNC scheme, each transceiver $\msT_i$ sends its message $B_i \in \{0,1\}$ to the relay through releasing $X_i=\zeta B_i$ molecules of type $\msM_i$. When both transceivers send the information bit $1$, almost all molecules react with each other and we have a physical-layer XOR. That is, the relay implicitly decodes the physically made XOR of the messages, $B_1 \oplus B_2$, and sends it to the transceivers through releasing $X_3$ molecules of type $\msM_3$.
%The block diagram of the system is shown in Fig. \ref{blockdiag}.
We define an auxiliary variable $B_{\msR_i}$ as the part of the message $B_1 \oplus B_2$ which corresponds to $B_i$.
Each receptor $i$ at the relay decodes the message $B_{\msR_i}=B_i \cdot (B_1 \oplus B_2)$, the part of the message $B_1 \oplus B_2$ which corresponds to $B_i$, as $\hat{B}_{\msR_i}$. For $\hat{B}_{\msR_1}=\hat{B}_{\msR_2}=0$, the relay stays silent; otherwise (when $\hat{B}_{\msR_1}=1$ or $\hat{B}_{\msR_2}=1$), it releases $\zeta_3$ molecules of type $\msM_3$ and hence, $X_3=(\hat{B}_{\msR_1}+\hat{B}_{\msR_2})\zeta_3=B_{\msR}\zeta_3$. Due to the perfect reaction assumption, $\hat{B}_{\msR_1}$ and $\hat{B}_{\msR_2}$ cannot be $1$ at the same time and thus, $X_3 \in \{0,\zeta_3\}$. We remark that these notations are used for the ease of error analysis.

In fact, the message sent by the relay ($B_{\msR}$) implicitly shows the $\widehat{B_1 \oplus B_2}$ and it is realized through $\hat{B}_{\msR_1}$ and $\hat{B}_{\msR_2}$ in our scheme. Furthermore, the system naturally adds up $\hat{B}_{\msR_1}$ and $\hat{B}_{\msR_2}$, because the encoder would release molecules when it is stimulated by the active receptor (at most one active receptor exists in each time slot).

The error events in this scheme are defined similar to the SNC scheme and the total error probability at the transceiver $\msT_i$ can be obtained from Lemma \ref{lemmaPei}.
Since the second communication phase is the same in both SNC and PNC schemes, $\mP(E^\msT_i|b_R)$ for both schemes can be obtained from Lemma \ref{lemmaPeTi}. In the following, we derive the error probability of the first phase.

In phase 1, when both transceivers send the same information bit $1$ or $0$, the concentrations of molecules of types $1$ and $2$ around the relay are $C_1=C_2=0$ (thanks to perfect reaction) and when the transceiver $\msT_i, i \in \{1,2\}$, sends the information bit $1$ and the transceiver $\msT_{\bar{i}}$ sends the information bit $0$, the concentrations are $C_i=\zeta \pi_1$ and $C_{\bar{i}}=0$. Hence, when $B_{\msR_i}=b_{\msR_i}$, the average concentration of the molecule type $\msM_i$ around the relay is $C_i=b_{\msR_i} \zeta\pi_1$, and due to the transparent receiver, the number of counted molecules of type $\msM_i$ at the relay, $Y_i^\msR$, is Poisson$(b_{\msR_i} \zeta\pi_1v_\textrm{r})$. Similar to the SNC scheme, the relay uses a threshold $\tau_i^{\msR}$, to decode $B_{\msR_i}$. According to Lemma \ref{lemmathreshold}, the optimum thresholds are obtained as $\tau_i^{\msR}=0, i \in \{1,2\}$.\footnote{Note that here we have, $\mP\{B_{\msR_i}=1\}=\mP\{B_i=1,B_{\bar{i}}=0\}=\frac{1}{4}$ and $\mP\{B_{\msR_i}=0\}=1-\mP\{B_{\msR_i}=1\}=\frac{3}{4}$. Hence, $\mP\{B_{\msR_i}=1\}<\mP\{B_{\msR_i}=0\}$ and the condition in Lemma \ref{lemmathreshold} is satisfied.}
Using these thresholds, the error probability of the first phase in the PNC scheme is obtained in Lemma \ref{lemmaPeiPNC}.
\begin{lemma}\label{lemmaPeiPNC}
The error probability of the first communication phase, $\mP(E_{\msR}|b_1,b_2)$, in the PNC scheme using the optimum threshold can be obtained as
\begin{align}\label{eq_PeRPNC_cond}
\mP(E_{\msR}|B_1=0,B_2=0)&=\mP(E_{\msR}|B_1=1,B_2=1)=0,\\\nonumber
\mP(E_{\msR}|B_1=1,B_2=0)&=\mP(E_{\msR}|B_1=0,B_2=1)=\exp(-\zeta\pi_1v_\textrm{r}).\end{align}
\end{lemma}
\begin{proof}
We define $E_{\msR_i}$ as the event $\{\hat{B}_{\msR_i} \neq B_{\msR_i}\}$. Hence, $\mP(E_{\msR_i})$ is the probability of error when $B_{\msR_i}$ is decoded with error at the $i$-th receptor of the relay. Since $\tau_i^\msR=0$, for $i \in \{1,2\}$,
\begin{align}\label{eq_PeRi}
%\begin{aligned}
\mP(E_{\msR_i}|B_{\msR_i}=0)=0, \quad \mP(E_{\msR_i}|B_{\msR_i}=1)=\exp(-\zeta\pi_1v_\textrm{r}).
%\end{aligned}
\end{align}
Recall that the number of released molecules of type $\msM_3$ equals to $X_3=0$ when the transceivers send the same messages and $X_3=\zeta_3$ when one of the transceivers send the information bit $1$ and the corresponding receptor at the relay decodes it correctly (see Table \ref{table1}).
\begin{table}
\centering
\caption{Messages and number of molecules sent by the relay in the proposed PNC scheme}
\begin{tabular}{|c|c|K{1.2cm}|K{3cm}|K{3cm}|c|}
\hline
$B_1$ & $B_2$ & $B_1 \oplus B_2 $ & $B_{\msR_1}=B_1 \cdot (B_1 \oplus B_2)$ & $B_{\msR_2}=B_2 \cdot (B_1 \oplus B_2)$ & $X_3$\\\hline\hline
$0$ & $0$ & $0$ & $0$ & $0$ & $0$\\
$1$ & $0$ & $1$ & $1$ & $0$ & $\hat{B}_{\msR_1} \zeta_3$\\
$0$ & $1$ & $1$ & $0$ & $1$ & $\hat{B}_{\msR_2} \zeta_3$\\
$1$ & $1$ & $0$ & $0$ & $0$ & $0$\\
\hline
\end{tabular}
\label{table1}
\vspace{-1.5em}
\end{table}
Thus, when $(B_1,B_2)\in\lbrace(0,0),(1,1)\rbrace$, $B_{\msR_1}$ and $B_{\msR_2}$ equal to zero and are decoded without error at the relay. When $(B_1,B_2)=(1,0)$, we have $B_{\msR_1}=1$ and $B_{\msR_2}=0$. Hence, $B_{\msR_2}$ is decoded without error at the relay and we get $\mP(E_{\msR}|B_1=1,B_2=0)=\mP(E_{\msR_1}|B_{\msR_1}=1)$. Similarly, we get $\mP(E_{\msR}|B_1=1,B_2=0)=\mP(E_{\msR_2}|B_{\msR_2}=1)$. %Therefore, we obtain \eqref{eq_PeRPNC_cond}.
%\begin{align}\label{eq_PeRPNC_cond}
%\mP(E_{\msR}|B_1=0,B_2=0)&=\mP(E_{\msR}|B_1=1,B_2=1)=0,\\\nonumber
%\mP(E_{\msR}|B_1=1,B_2=0)&=\mP(E_{\msR}|B_1=0,B_2=1)=\exp(-\zeta\pi_1v_\textrm{r}).
%\end{align}
%Now, by taking average over $B_1$ and $B_2$, we obtain \eqref{eq_PeRPNC}.
\end{proof}
Now, by substituting the error probabilities of the two phases from \eqref{eq_PeRPNC_cond} and \eqref{eq_PeRihat} in \eqref{eq_Peim1m2}, we obtain
\begin{align}\label{Peireaction1}
p_{\textrm{e},i}=&\frac{1}{2}\exp(-\zeta_3\pi_1v_\textrm{r})+\frac{1}{2}\left[1-\exp(-\zeta_3\pi_1v_\textrm{r})\right]\exp(-\zeta\pi_1v_\textrm{r}),
\end{align}
for $i \in \{1,2\}$, and thus the Avg-BEP can be obtained from \eqref{AvgBEP}.

\begin{remark} Comparing \eqref{Peitype1} and \eqref{Peireaction1}, it can be seen that the error probability at each transceiver and
thus the Avg-BEP of the PNC is lower than the SNC since $\exp(-\zeta\pi_1v_{\textrm{r}})>\exp(-2\zeta\pi_1v_{\textrm{r}})$ and hence, $\exp(-\zeta\pi_1v_{\textrm{r}})<2\exp(-\zeta\pi_1v_{\textrm{r}})-\exp(-2\zeta\pi_1v_{\textrm{r}})$. In fact, since $\mP(E_\msR|B_1=1,B_2=1)$ is zero for the PNC scheme, but it is positive for the SNC scheme, the error probability at the relay is lower for the PNC scheme. Further, $\mP(E^{\msT_i}|B_{\msR}=0)$ and $\mP(E^{\msT_i}|B_{\msR}=1)$ are equal for both schemes. Thus, according to \eqref{eq_Peim1m2}, the error probability at the transceivers is lower for the PNC scheme. 
%The error probability is derived in \eqref{eq_Pei} as $p_{\textrm{e},i}=\mP(E^{\msT_i})+\big(1-2\mP(E^{\msT_i})\big)\mP(E_{\msR})$;
%First, note that $\mP(E_\msR)$ in the PNC scheme is lower than the SNC scheme (because $\mP(E_\msR|B_1=1,B_2=1)$ is zero for the PNC scheme, but it is positive for the SNC scheme); Second, $\mP(E^{\msT_i}|B_{\msR}=0)$ and $\mP(E^{\msT_i}|B_{\msR}=1)$ are equal for both schemes.
\end{remark}

\vspace{-1em}
\section{Error Performance Analysis in the Presence of ISI}\label{errorISI}
For simplicity of exposition, we assume the transceiver-relay and the relay-transceiver channels to have unit super time slot memory.
In Section \ref{Simulation}, we simulate the system for higher channels memories.

\subsection{The SNC scheme}
Similar to the no ISI case, from \eqref{eq_Peim1m2}, we define two error events in each super time slot corresponding to each communication phase: (i) $E_{\msR,k}=\{\hat{B}_{\msR,k} \neq B_{\msR,k}\}$, and (ii) $E_k^{\msT_i}=\{\hat{B}_{\msR,k}^{\msT_i} \neq \hat{B}_{\msR,k}\}$. In the following, we obtain the error probabilities of both communication phases.

\textbf{Phase 1:} According to \eqref{XikSNCbargashti}, the transceiver $\msT_i$ uses the number of released molecules in the $(k-1)$-th super time slot to determine the number of released molecules in the $k$-th super time slot: if its message is 1, the transceiver $\msT_i$ releases some molecules such that the concentration of molecules of type $\msM_i$ at the relay will be equal to $c_\textrm{r}$ and if its message is 0, it stays silent (concentration of the molecules of type $\msM_i$ at the relay will be equal to the concentration of the remained molecules from the previous super time slot, i.e., $X_{i,k-1} \pi_3$). Hence, given $B_{i,k}=b_{i}$ and $X_{i,k-1}=x$, the average concentration of molecules of type $\msM_i$ around the relay is $C_{i,k}=b_{i} c_\textrm{r}+(1-b_{i})x\pi_3$ and the number of counted molecules of type $\msM_i$ at the relay, $Y_{i,k}^\msR$, has a Poisson distribution with parameter $C_{i,k}v_\textrm{r}$. It is just straightforward to show from \eqref{XikSNCbargashti} that the probability distribution function (PDF) of $X_{i,k}$ for $i \in \{1,2\}$ is as follows:
\begin{align}\label{DistXi}
p_{X_{i,k}}(x)=\sum_{n=1}^{\infty}(\frac{1}{2})^{n} \delta(x-x_{n}), \qquad x_{n}=\frac{c_\textrm{r}}{\pi_1}\sum_{l=0}^{n-2}(-\eta_3)^l, \qquad n\in \mathbb{N}.
\end{align}
The relay uses MAP decision rule to decode the message of the transceiver $\msT_i, i \in \{1,2\}$, \emph{i.e.}, $B_{i,k}$, as
\begin{align}
\frac{1}{2}\mP(y_{i,k}^{\msR}|B_{i,k}=1) \overset{1}{\underset{0}\gtrless} \frac{1}{2}\mP(y_{i,k}^{\msR}|B_{i,k}=0).
\end{align}
Noting that $y_{i,k}^\msR$ conditioned on $B_{i,k}=0$ depends on $X_{i,k-1}$, and taking average of $\mP(y_{i,k}^{\msR}|B_{i,k}=0)$ over $X_{i,k-1}$ results in
\begin{equation}
\begin{aligned}
\mP(y_{i,k}^{\msR}|B_{i,k}=1)\overset{1}{\underset{0}\gtrless}\int\limits_{0}^{\infty} p_{X_{i,k-1}}(x)\mP(y_{i,k}^{\msR}|B_{i,k}=0,X_{i,k-1}=x) dx.
\end{aligned}
\end{equation}
By substituting $p_{X_{i,k-1}}(x)$ from \eqref{DistXi}, we obtain the MAP decision rule at the $i$-th receptor as

\begin{equation}\label{eq:decSNC}
\begin{aligned}
%\mP(y_{i,k}^{\msR}|B_{i,k}=1)-\sum_{m=1}^{\infty} (\frac{1}{2})^{m} \mP(y_{i,k}^{\msR}|B_{i,k}=0,X_{i,k-1}=x_m)\overset{1}{\underset{0}\gtrless}0.
%\\
(c_\textrm{r}v_\textrm{r})^{y_{i,k}^{\msR}}\exp(-c_\textrm{r}v_\textrm{r}) \overset{1}{\underset{0}\gtrless} \sum_{n=1}^{\infty} (\frac{1}{2})^{n} (x_n\pi_3v_\textrm{r})^{y_{i,k}^{\msR}}\exp(-x_n\pi_3v_\textrm{r}).
\end{aligned}
\end{equation}
It can be shown that the above decision rule yields to a simple threshold rule, i.e., $y_{i,k}^{\msR} \overset{1}{\underset{0}\gtrless} \tau_i^{\msR}$, where the optimum threshold $\tau_i^{\msR}$ can be found numerically.
% from \eqref{eq:decSNC}. 
Then, the error probability at the $i$-th receptor ($i \in \{1,2\}$) of the relay is obtained as
\begin{align}\label{eq_PeRISI}
\mP(E_{i,k}^{\msR}|B_{i,k}=0)&=\sum_{n=1}^{\infty}(\frac{1}{2})^{n}\mP\{Y_{i,k}^\msR> \tau_i^{\msR}|B_{i,k}=0,X_{i,k-1}=x_n\},\\\nonumber
\mP(E_{i,k}^{\msR}|B_{i,k}=1)&=\mP\{Y_{i,k}^\msR\leq \tau_i^{\msR}|B_{i,k}=1\},
\end{align}
Now, $\mP(E_{\msR,k}|B_{1,k}=b_1,B_{2,k}=b_2)$ can be obtained by substituting \eqref{eq_PeRISI} in \eqref{eq_PeR}.

\textbf{Phase 2:} Here, using fixed transmission rate, the number of counted molecules at $\msT_i$ when $B_{\msR,k}=b_{\msR}$ and $B_{\msR,k-1}=\tilde{b}_{\msR}$) is $Y_{3,k}^{\msT_i}\sim$Poisson$(b_{\msR} \zeta_3 \pi_1v_\textrm{r}+\tilde{b}_{\msR} \zeta_3 \pi_3v_\textrm{r})$. 
To mitigate ISI in this phase, the transceiver $\msT_i$ uses the decoded message of the relay in the $(k-1)$-th super time slot, i.e., $\hat{B}_{\msR,k-1}^{\msT_i}$ and obtains the adaptive threshold in the $k$-th super time slot using Maximum Likelihood (ML) decision rule. The adaptive threshold at the transceiver $\msT_i$ is obtained in Lemma \ref{lemma_adapt} for the relay-transceiver channel with one super time slot memory.
\begin{lemma}\label{lemma_adapt}For a relay-transceiver channel with one super time slot memory, the adaptive threshold at the transceiver $\msT_i$ to decode the message of the relay, using its decoded message in the previous super time slot ($\hat{B}_{\msR,k-1}^{\msT_i}=\hat{b}_{\msR}$), is obtained according to ML decision rule as
\begin{align}
\tau^{\msT_i}(\hat{b}_{\msR})=\frac{\zeta_3 \pi_1v_\textrm{r}}{\ln(1+\frac{1}{\eta_3\hat{b}_{\msR}})}.
\end{align}
\end{lemma}
\begin{proof}
Using ML decision rule to decode $B_{\msR,k}$ at the transceiver $\msT_i$, we have
\begin{align}
\mP(y_{3,k}^{\msT_i}|B_{\msR,k}=1, B_{\msR,k-1}=\hat{b}_{\msR})\overset{1}{\underset{0}\gtrless}\mP(y_{3,k}^{\msT_i}|B_{\msR,k}=0,B_{\msR,k-1}=\hat{b}_{\msR}),
\end{align}
Hence,
\begin{equation}
\begin{aligned}
y_{3,k}^{\msT_i}\overset{1}{\underset{0}\gtrless}\frac{\zeta_3 \pi_1v_\textrm{r}}{\ln(1+\frac{1}{\eta_3\hat{b}_{\msR}})}=\tau^{\msT_i}(\hat{b}_{\msR}).
\end{aligned}
\end{equation}
which gives the adaptive threshold used at $\msT_i$ (that is $\tau^{\msT_i}(\hat{b}_{\msR})$). \end{proof}
It can be easily seen that when previous decoded message is zero, our ISI mitigating technique gives the zero threshold (i.e., $\tau^{\msT_i}(0)=0$).
For the above decision rule, the error probability at $\msT_i$ for $b_{\msR}\in\{0,1\}$ is obtained as
\begin{align}\label{eq_PeRihatISI}
\mP(E_k^{\msT_i}|B_{\msR,k}=b_{\msR})=\sum_{\tilde{b}_{\msR},\hat{b}_{\msR} \in \{0,1\}} &\left[ \mP\{B_{\msR,k-1}=\tilde{b}_{\msR}\} \mP\{\hat{B}_{\msR,k-1}^{\msT_i}=\hat{b}_{\msR}|\tilde{b}_{\msR}\} \mP\big(E_k^{\msT_i}|b_{\msR}, \tilde{b}_{\msR},\hat{b}_{\msR}\big)\right],
\end{align}
for $ i \in \{1,2\}$, where 
\begin{align}\nonumber
&\mP(B_{\msR,k-1}=0)=\frac{1}{4}\big[2-\mP(E_{\msR,k-1}|B_{1,k-1}=0,B_{2,k-1}=0)-\mP(E_{\msR,k-1}|B_{1,k-1}=1,B_{2,k-1}=1)\\\label{eqpBR}
&\qquad \qquad \qquad \quad+\mP(E_{\msR,k-1}|B_{1,k-1}=0,B_{2,k-1}=1)+\mP(E_{\msR,k-1}|B_{1,k-1}=1,B_{2,k-1}=0)\big],\\
&\mP(\hat{B}_{\msR,k-1}^{\msT_i}=\hat{b}_{\msR}|\tilde{b}_{\msR})=\begin{cases}
\mP(E_{k-1}^{\msT_i}|\tilde{b}_{\msR}), \quad &\textrm{if } \hat{b}_{\msR} \neq \tilde{b}_{\msR},\\
1-\mP(E_{k-1}^{\msT_i}|\tilde{b}_{\msR}), & \textrm{if } \hat{b}_{\msR}= \tilde{b}_{\msR},\end{cases}
\end{align}
and $\mP\big(E_k^{\msT_i}|B_{\msR,k}=0, \tilde{b}_{\msR},\hat{b}_{\msR}\big)=\mP\big\{Y_{3,k}^{\msT_i}>\tau^{\msT_i}(\hat{b}_{\msR})|B_{\msR,k}=0, \tilde{b}_{\msR}\big\}$, $\mP\big(E_k^{\msT_i}|B_{\msR,k}=1, \tilde{b}_{\msR},\hat{b}_{\msR}\big)=\mP\big\{Y_{3,k}^{\msT_i}\leq\tau^{\msT_i}(\hat{b}_{\msR})|B_{\msR,k}=1, \tilde{b}_{\msR}\big\}$. Hence, $\mP(E_k^{\msT_i}|B_{\msR,k}=b_{\msR})$ can be obtained recursively from \eqref{eq_PeRihatISI}. Since we have two linear equations in \eqref{eq_PeRihatISI} with two unknowns, a closed form equation can be easily obtained for $\mP(E_k^{\msT_i}|B_{\msR,k}=b_{\msR})$.

\begin{remark} (No error (NoE) approximation) The total error probability at the transceiver $\msT_i$ can be further simplified assuming that the message of the relay is decoded without error at the transceivers in the previous super time slot (i.e., we ignore the error propagation). Hence, the error probability of the second phase can be obtained simply from \eqref{eq_PeRihatISI} assuming $\mP(E_{k-1}^{\msT_i}|\tilde{b}_\msR)=0$, and the total error probability at the transceiver $\msT_i$ can be obtained for $ i \in \{1,2\}$ using \eqref{eq_Pei}: %(see Appendix \ref{ProoflemmalowerPNC}):
\begin{equation}
\begin{aligned}
\label{eq_PeiNoESNC}
p_{\textrm{e},i}^\textrm{NoE}=\frac{1}{16}\big((2-u_1)^2-u_2^2\big) w_{i,1}+\frac{1}{16}\big((2-u_2)^2-u_1^2\big) w_{i,2}+\frac{1}{4}(u_1+u_2),
\end{aligned}
\end{equation}
where $u_1=\mP(E_{\msR,k}|B_{1,k}=0,B_{2,k}=0)+\mP(E_{\msR,k}|B_{1,k}=1,B_{2,k}=1)$, $u_2=\mP(E_{\msR,k}|B_{1,k}=0,B_{2,k}=1)+\mP(E_{\msR,k}|B_{1,k}=1,B_{2,k}=0)$ can be computed from \eqref{eq_PeRISI} and \eqref{eq_PeR}, and
\begin{align}\label{eq_E}
w_{i,1}&=\exp(-\zeta_3\pi_1v_\textrm{r})+\sum_{l=\tau^{\msT_i}(1)+1}^{\infty}\frac{(\zeta_3\pi_3v_\textrm{r})^l}{l!}\exp(-\zeta_3\pi_3v_\textrm{r}),\\\nonumber
w_{i,2}&= \sum_{l=0}^{\tau^{\msT_i}(1)}\frac{(\zeta_3(\pi_1+\pi_3)v_\textrm{r})^l}{l!}\exp(-\zeta_3(\pi_1+\pi_3)v_\textrm{r}).
\end{align}
This provides a lower bound on the error probability of the SNC scheme as follows: from \eqref{eq_Peim1m2}, we have $p_{\textrm{e},i}=\mP(E_{\msR})+\frac{1}{2}\big(1-2\mP(E_{\msR})\big)[\mP(E^{\msT_i}|B_\msR=0)+\mP(E^{\msT_i}|B_\msR=1)]$; by ignoring the error propagation, we obtain lower bounds on $\mP(E^{\msT_i}|B_\msR=0)$ and $\mP(E^{\msT_i}|B_\msR=1)$; now, since $\mP(E_{\msR})\leq0.5$, this is a lower bound on $p_{\textrm{e},i}$.
\end{remark}

%\vspace{-0.5em}
\subsection{The PNC scheme}\label{errorISI:PNC}
Here, the error probability of the second phase is the same as that of the SNC given in \eqref{eq_PeRihatISI}, with the difference that $\mP(B_{\msR,k-1}=0)$ must be computed separately from \eqref{eqpBR}, since the error probabilities of the first phase are not equal for two schemes. Thus, we only analyze the error probability of the first phase.
According to \eqref{XikPNCbargashti}, the transceiver $\msT_i$ uses the decoded message of the other transceiver in the $(k-1)$-th super time slot ($\hat{B}_{\bar{i},k-1}^{\msT_i}$) and the number of its own released molecules in the $(k-2)$-th super time slot ($X_{i,k-2}$) to determine the number of released molecules in the $k$-th super time slot.
$X_{1,k-2}$ and $X_{2,k-2}$, themselves, depend on the previous decoded messages and hence, they may contain error. We consider the error effect in $(k-1)$-th super time slot and neglect the error effect in prior super times slots %(i.e., $X_{i,k-2}$) 
to obtain an approximate value for the error probability of the first communication phase (however, in Section \ref{Simulation}, we simulate this system and obtain the error probability considering the effect of error in $X_{i,k-2}$). With this assumption, the error probability of phase 1 in the $k$-th super time slot is obtained as
\begin{align}\label{eq_peRPNCISI}
&\mP(E_{\msR,k}|B_{1,k}=b_1,B_{2,k}=b_2)=\frac{1}{4}\sum_{\substack{\hat{b}_1, \hat{b}_2,\tilde{b}_1,\tilde{b}_2 \in \{0,1\}}} \left[\mP\{\hat{B}_{1,k-1}^{\msT_2}=\hat{b}_1,\hat{B}_{2,k-1}^{\msT_1}=\hat{b}_2|b_1,b_2,\tilde{b}_1,\tilde{b}_2\} \right.\\\nonumber
&\qquad\qquad \qquad \left. \times \mP\big(E_{\msR,k}|(B_{1,k},B_{2,k}, B_{1,k-1}, B_{2,k-1},\hat{B}_{1,k-1}^{\msT_2}, \hat{B}_{2,k-1}^{\msT_1})=(b_1,b_2, \tilde{b}_1, \tilde{b}_2,\hat{b}_1,\hat{b}_2)\big)\right].
\end{align}
The first term in the summation of \eqref{eq_peRPNCISI} is the joint decoding probability at the transceivers, which is independent of the current messages ($b_1,b_2$) and can be derived as a function of the error probabilities in the $(k-1)$-th super time slot as
\begin{align}\label{eq_state}\nonumber
\mP\{\hat{B}_{1,k-1}^{\msT_2}=\hat{b}_1,&\hat{B}_{2,k-1}^{\msT_1}=\hat{b}_2|b_1,b_2,\tilde{b}_1,\tilde{b}_2\}=\mP\{\hat{B}_{1,k-1}^{\msT_2}=\hat{b}_1,\hat{B}_{2,k-1}^{\msT_1}=\hat{b}_2|\tilde{b}_1,\tilde{b}_2\}\\\nonumber
&=\big(1-\mP(E_{\msR,k-1}|\tilde{b}_1,\tilde{b}_2)\big)\mP\{\hat{B}_{1,k-1}^{\msT_2}=\hat{b}_1,\hat{B}_{2,k-1}^{\msT_1}=\hat{b}_2|\tilde{b}_1,\tilde{b}_2,E_{\msR,k-1}^{c}\}\\
&\qquad \qquad+\mP(E_{\msR,k-1}|\tilde{b}_1,\tilde{b}_2) \mP\{\hat{B}_{1,k-1}^{\msT_2}=\hat{b}_1,\hat{B}_{2,k-1}^{\msT_1}=\hat{b}_2|\tilde{b}_1,\tilde{b}_2,E_{\msR,k-1}\}.
\end{align}
Now, considering the independent decoding at the transceivers, as well as the independent channels from the relay to the transceivers, we obtain
\begin{align}\label{eq_state2}
\mP\{\hat{B}_{1,k-1}^{\msT_2}=\hat{b}_1&,\hat{B}_{2,k-1}^{\msT_1}=\hat{b}_2|\tilde{b}_1,\tilde{b}_2,E_{\msR,k-1}^{c}\}\\\nonumber
&=\mP\{\hat{B}_{1,k-1}^{\msT_2}=\hat{b}_1|\tilde{b}_1,\tilde{b}_2,E_{\msR,k-1}^{c}\} \mP\{\hat{B}_{2,k-1}^{\msT_1}=\hat{b}_2|\tilde{b}_1,\tilde{b}_2,E_{\msR,k-1}^{c}\},
\end{align}
where the above probabilities would be the error probability in decoding the message of the relay at $T_i$ when $\hat{b}_i \neq \tilde{b}_i$, for $i \in \{1,2\}$, and thus:
\begin{equation}\label{eq_state3}
\begin{aligned}
\mP\{\hat{B}_{i,k-1}^{\msT_{\bar{i}}}&=\hat{b}_i|\tilde{b}_1,\tilde{b}_2,E_{\msR,k-1}^{c}\}=\begin{cases}
\mP(E_{k-1}^{\msT_i}|B_{\msR,k-1}=\tilde{b}_1\oplus \tilde{b}_2), &\textrm{if }\hat{b}_i \neq \tilde{b}_i\\
1-\mP(E_{k-1}^{\msT_i}|B_{\msR,k-1}=\tilde{b}_1\oplus \tilde{b}_2), &\textrm{if } \hat{b}_i= \tilde{b}_i
\end{cases}.
\end{aligned}
\end{equation}
Similar equations can be derived for $\mP\{\hat{B}_{1,k-1}^{\msT_2}=\hat{b}_1,\hat{B}_{2,k-1}^{\msT_1}=\hat{b}_2|\tilde{b}_1,\tilde{b}_2,E_{\msR,k-1}\}$.
Combining \eqref{eq_state}-\eqref{eq_state3} gives the first term in the summation of \eqref{eq_peRPNCISI}.
To obtain the second term in the summation of \eqref{eq_peRPNCISI}, i.e., $\mP\big(E_{\msR,k}|(B_{1,k},B_{2,k}, B_{1,k-1}, B_{2,k-1},\hat{B}_{1,k-1}^{\msT_2},$ $ \hat{B}_{2,k-1}^{\msT_1})=(b_1,b_2, \tilde{b}_1, \tilde{b}_2,\hat{b}_1,\hat{b}_2)\big)$, one must obtain the concentration of each molecule type around the relay after reaction for all $2^6$ realizations of $b_1,b_2, \tilde{b}_1, \tilde{b}_2, \hat{b}_1, \hat{b}_2$. Then, the error probability at the relay for each case can be derived based on the corresponding average number of counted molecules. The details are given in Appendix~\ref{apendixa}, where the second term in the summation of \eqref{eq_peRPNCISI} is derived. Combing all these equations, a set of recursive equations is obtained for the error probability of the relay in Appendix~\ref{apendixa}.

\begin{remark} (NoE approximation) To further simplify the error performance results, we consider the case where there is no error in the decoded messages of the previous super time slots. Then, the error probability of phase 1 will be equal to the no ISI case, and the error probability of the second phase can be obtained similar to the SNC scheme.
%For the error probability of phase 2, we take the average of \eqref{eq_PeRihatISI} over $\hat{B}_{\msR}^{\msT_i,k-1}=B_{\msR,k-1}$ and
Hence, we obtain the total error probability at the transceiver $\msT_i$ as follows:
 %(see Appendix \ref{ProoflemmalowerPNC} for details):
%. This lower bound is computed in Lemma \ref{lemmalowerPNC}.
%\begin{lemma}\label{lemmalowerPNC}
%A lower bound on the error probability at the transceiver $\msT_i$ in the PNC scheme for $i \in \{1,2\}$ is obtained as follows:
\begin{align}
\label{eq_PeiNoEPNC}
p_{\textrm{e},i}^\textrm{NoE}=&\frac{1}{16}(4-u^2) w_{i,1}+\frac{1}{16}(2-u)^2 w_{i,2}+\frac{1}{4}u,
\end{align}
for $i \in \{1,2\}$, where
$u=\exp(-\zeta_1\pi_1v_\textrm{r})+\exp(-\zeta_2\pi_1v_\textrm{r})$, and $w_{i,1}$, $w_{i,2}$ are defined in \eqref{eq_E}.
Note that, ignoring the error propagation gives lower bounds on the error probabilities of each hop, while the overall error probability cannot be proved to necessarily be a lower bound. However, in our simulation results, it is always a lower bound.
%However, the numerical and simulation results in Section \ref{sec:sim} show that in practice it is a lower bound.
%\vspace{-0.5em}
\end{remark}

\section{Simulation and Numerical Results}\label{Simulation}
In this section, we evaluate the performance of the PNC and SNC schemes in terms of the probability of error. We consider the parameters $D=10^{-9}~\textrm{m}^2/\textrm{s}$, $d=250~ \textrm{nm}$, and $v_\textrm{r}=\frac{4}{3}\pi (50 )^3~\textrm{nm}^3$ (consistent with prior works \cite{IWCIT2016type, Jamali2017} ). In the no ISI case, we choose $t_s=t_0=10.4167~\mu\textrm{s}$ which is the time that the impulse responses of the channels take their maximum. In the ISI case, we assume $t_0=10.4167~\mu\textrm{s}$ and $t_s$ is chosen such that $\eta_{q+2}=0.05$. For the simulations, we run the Monte-Carlo simulations for $5 \times 10^6$ transmitted bits (based on uniform distribution). For each bit, we generate a random number using the distribution of the number of received molecules. Then, using the derived threshold in the receiver, we decode each transmitted bit, and hereby estimate the bit error probability with \emph{bit error rate}.

\begin{figure}
\centering
\begin{minipage}{.43\textwidth}
\centering
\vspace{0pt}
\includegraphics[trim={1cm 0 0 0cm}, scale=0.38]{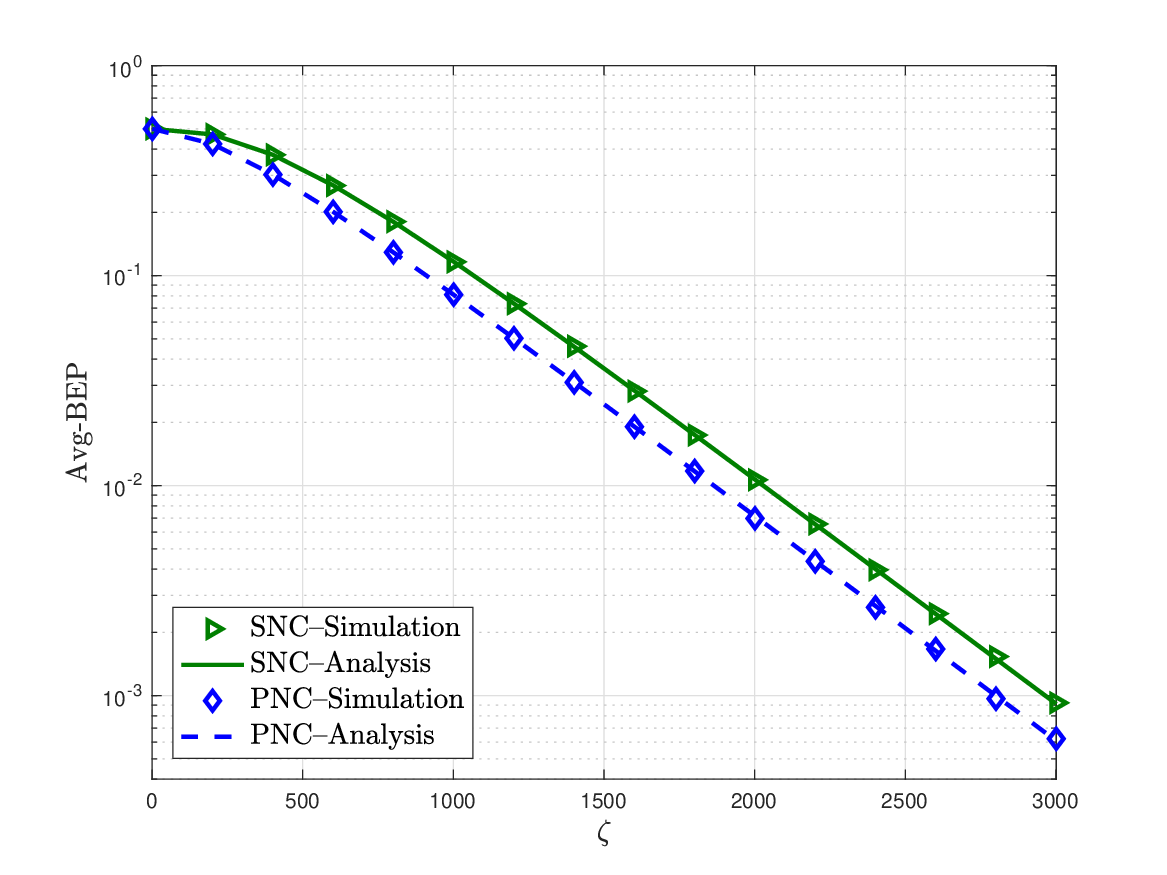}
\caption{Average bit error probability with respect to $\zeta_1=\zeta_2=\zeta_3=\zeta$ without ISI.}
\label{fignoISI}
\end{minipage}\qquad%
\begin{minipage}{0.47\textwidth}
\centering
\includegraphics[trim={1cm 0 0 -0.5cm}, scale=0.38]{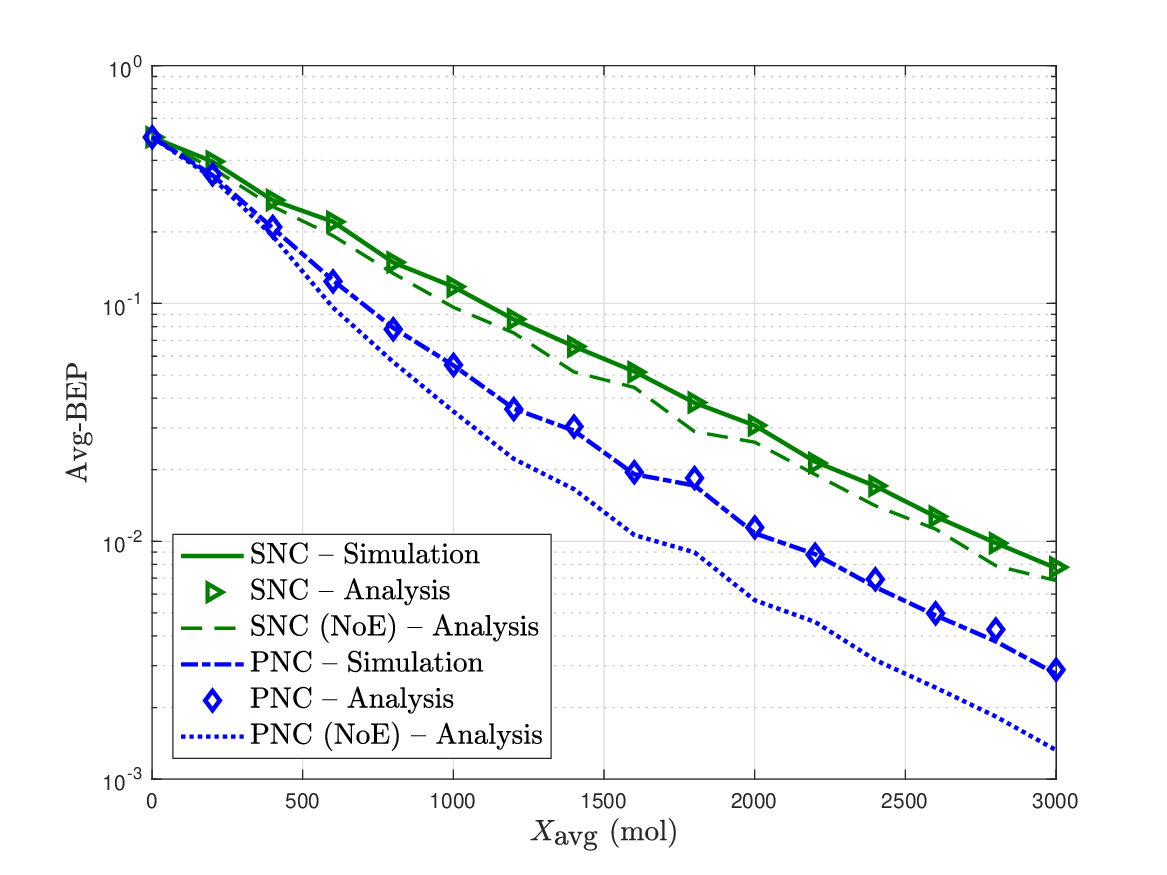}
\caption{Avg-BEP with respect to the average number of transmitted molecules in the presence of ISI ($q=3$).}
\label{figISI}
\vspace{-1.5em}
\end{minipage}
\end{figure}

Fig. \ref{fignoISI} shows the Avg-BEP versus the number of transmitted molecules from the transceivers and the relay for information bit 1 (i.e., $\zeta_1=\zeta_2=\zeta_3=\zeta$) for the two schemes without ISI using \eqref{AvgBEP}, \eqref{Peireaction1}, and \eqref{Peitype1} along with the Avg-BEP using simulation. It can be seen that the proposed PNC scheme outperforms the SNC scheme. This is due to the reduction in the number of the molecules bound to the receptors (thanks to reaction). Also, the simulations confirm the analytical results.

Fig. \ref{figISI} shows the Avg-BEP versus the average number of transmitted molecules from the transceivers and the relay (i.e., $\frac{1}{2}\sum_{i=1}^{2}X_{i,\textrm{avg}}^{\textrm{PNC}}=\frac{1}{2}\sum_{i=1}^{2} X_{i,\textrm{avg}}^{\textrm{SNC}}=\frac{1}{2}\zeta_3=X_\textrm{avg}$) in the presence of ISI using analysis and simulation along with the Avg-BEP using NoE approximations given in \eqref{eq_PeiNoEPNC} and \eqref{eq_PeiNoESNC} for the channels with memory of $3$. It can be seen that the error performance of the SNC scheme, for which we adopt the existing ISI mitigating techniques, is considerably worse than the error performance of the PNC scheme, for which we have proposed a reaction-based ISI mitigating technique.
The reason is that in the SNC scheme, using adaptive rate at each transceiver mitigates the ISI only when the message of the transceiver is $1$. But, in the PNC scheme, using adaptive rates at the transceivers mitigates the ISI in all cases of the sent messages. It is also seen that the NoE approximation of error probability of the PNC scheme is a lower bound.

\begin{figure}
\centering
\begin{minipage}{.43\textwidth}
\centering
\vspace{0pt}
\includegraphics[trim={1cm 0 0 2cm}, scale=0.38]{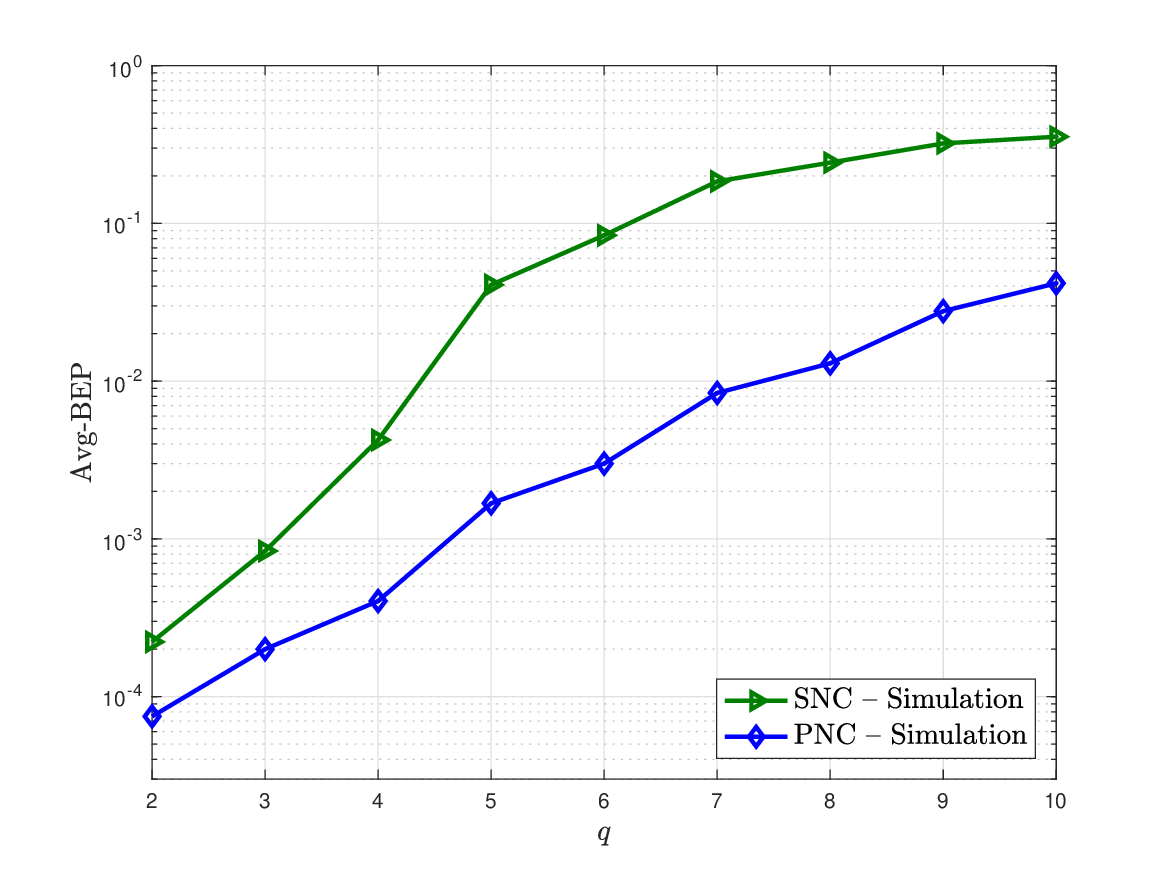}
\caption{Avg-BEP with respect to channel memory ($q$), in the presence of ISI ($X_{\textrm{avg}}=5000$).}
\label{figISIperq}
\end{minipage}\qquad%
\begin{minipage}{0.47\textwidth}
\centering
\includegraphics[trim={1cm 0 0 0}, scale=0.38]{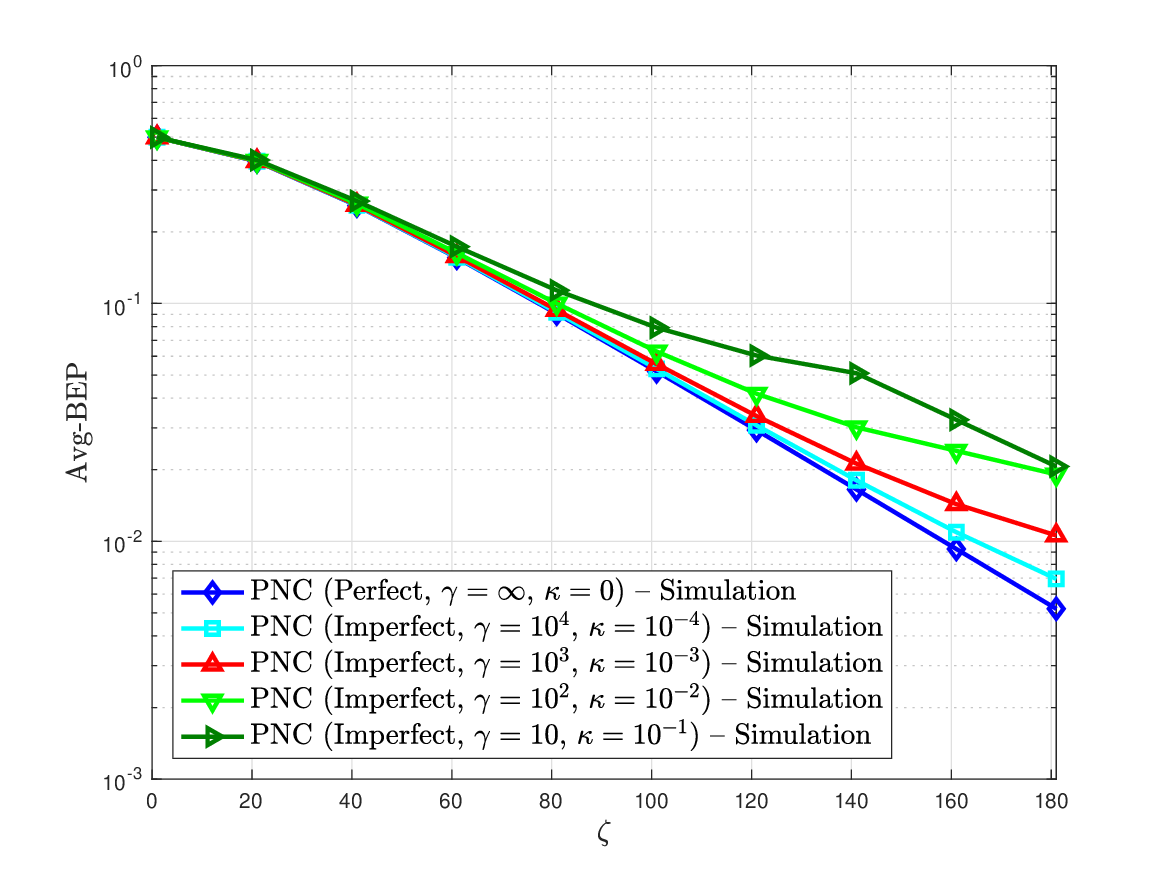}
\caption{Average bit error probability of the PNC scheme in 1-D with respect to $\zeta_1=\zeta_2=\zeta_3=\zeta$ without ISI for perfect and imperfect reaction among molecules.
% ($\gamma=0.01,0.1,1, \infty~\textrm{Molecule}^{-1}. \textrm{m}. \textrm{s}^{-1}$)
}
\label{figNoISIImperfect}
\end{minipage}
\vspace{-1.5em}
\end{figure}

Fig. \ref{figISIperq} shows the Avg-BEP versus the channel memory ($q$), in the presence of ISI. Here, we assume $\frac{1}{2}\sum_{i=1}^{2}X_{i,\textrm{avg}}^{\textrm{PNC}}=\frac{1}{2}\sum_{i=1}^{2} X_{i,\textrm{avg}}^{\textrm{SNC}}=5000$. It can be observed that the error probability increases by channel memory. However, the PNC scheme using the proposed ISI mitigating technique performs much better than the SNC scheme using the existing ISI mitigating techniques.

In Fig. \ref{figNoISIImperfect}, we consider the effect of imperfect reaction on the error performance of the PNC scheme. Fig. \ref{figNoISIImperfect} depicts the Avg-BEP of the PNC scheme without ISI versus $\zeta_1=\zeta_2=\zeta_3=\zeta$, when we have imperfect reaction among molecules of type $\msM_1$ and $\msM_2$. More specifically, we assume a finite forward reaction rate constant and a positive reverse reaction rate constant among molecules ($\gamma=10,10^2, 10^3,10^4~\textrm{Molecule}^{-1}. \textrm{m}. \textrm{s}^{-1}$ and $\kappa=10^{-1},10^{-2},10^{-3},10^{-4}~ \textrm{s}^{-1}$, respectively) and solve the reaction-diffusion equations in \eqref{eqreact} for 1-D environment numerically using the finite difference method (FDM) to obtain the concentration of each molecule type at the relay\cite{Leveque2007}.\footnote{Note that because of the time complexity of the FDM for 3-D environment, to see the effect of imperfect reaction on the PNC scheme, we simulated the system for 1-D instead of 3-D} Using the FDM, we discretize time and space to small sections and approximate derivations with difference equations in each section. Here, we assumed $v_\textrm{r}=30~\textrm{nm}$ and $t_s=31.25~\mu \textrm{s}$. The optimum thresholds are not zero for imperfect reaction and are obtained using simulation such that the error probability is minimized. It is observed from this figure that for the considered parameters, by increasing the forward reaction rate constant and decreasing the reverse reaction rate constant, the average error probability gets closer to that of the PNC with perfect reaction.
%and for the $\gamma=10^4$ and $\kappa=10^{-4}$, the system performs almost like the perfect reaction case.

%\section{Future Works}\label{sec:futurework}
% It seems to us that the existing approach may benefit from the following ideas:
%\begin{itemize}
%\item Instead of just looking at qualitative behavior, one can linearize the non-linear system around a steady
%state point of operation (just as in circuit theory where we linearize non-linear transistors by first finding the operating dc point, and then linearize the system and look at small variation around the dc point). The limitation of this technique is that it only allows utilizing small perturbation for modulation of information.
%\item One may utilize the rich mathematical theory of reaction--diffusion systems. For instance, it is known that some non-linear reaction-diffusion systems exhibit bifurcations in which a small change in a system parameter causes a sudden qualitative change in its behavior \cite{nicolis1974dissipative, auchmuty1976bifurcation, auchmuty1975bifurcation}. Furthermore, much is known in the math literature about the qualitative behavior of the solutions, \emph{e.g.},
%the existence of solutions of such systems, the boundedness, stability and asymptotics of their solutions, rates of convergence, geometry and topology of attracting sets, appearance of travelling wave solutions, etc (e.g., see \cite{Rothe, Murray}).
%\end{itemize}

%\vspace{-0.5em}
\section{Concluding Remarks}\label{conclusion}
In this paper, we proposed the physical-layer network coding (PNC) for molecular communication (MC) called the reaction-based PNC scheme, where we used different molecule types, reacting with each other by a fast reversible  reaction. Hence, we constructed a physical-layer XOR in this scheme without requiring an XOR gate at the relay. This results in a simple implementation for the proposed scheme compared to the straightforward network coding (SNC) scheme. To mitigate the ISI, we also used the reaction characteristics of the PNC scheme and proposed a reaction-based ISI mitigating technique for this scheme, where each transceiver using its previously decoded messages, cancels out the ISI of the other transceiver using the reaction of molecules. Considering the transparent receivers with the signal dependent noise, we investigated the error probabilities of the straightforward and the proposed network coding schemes. As expected and confirmed by simulations, the reaction-based scheme decreases the overall error probability in two-way relay MC, while having less complexity. Further, the proposed ISI mitigating technique for the PNC scheme has significantly better performance compared to the SNC scheme in which the existing techniques are applied to each hop. The main reason is that in the SNC, using adaptive transmission rate at each transceiver, mitigates its own ISI only when its message is 1. However, in the PNC, using adaptive rates at the transceivers mitigates the ISI for all sent messages. The effect of the imperfect reaction on the error probability of the PNC scheme is considered in simulations using the finite difference method (FDM). 

%\emph{Channel state information (CSI)}:
We have assumed that channel state information (CSI) is known at the transceivers, i.e., the transceivers know the channel coefficients of both transceivers to the relay channels. This is justified if the nodes have fixed distance, where the channel coefficients can be computed from the diffusion equation. Studying the network coding schemes with limited (or no) CSI is an interesting future work.
%\emph{Deterministic model}: 
Further, we considered the deterministic model for our analysis which ignores the channel noise. While in the presence of noise, the derivations would be much more complex, the methods do not change.

%\vspace{-1em}
%\section{Acknowledge}\label{Acknowledge}
%We would like to thank Dr. Atefeh Amerizadeh, for her helpful comments in possibility of molecular reactions.

\bibliographystyle{ieeetr}
\bibliography{reftest}

\appendices
\section{Error Probability of Phase 1 in the Proposed PNC Scheme in the Presence of ISI}\label{apendixa}
%Let $C_{i,k}$ be the concentration of molecules of type $\msM_i$ around the relay after reaction.
Consider the set of all 16 cases of the sent and decoded messages of the transceivers in the previous super time slot (i.e., the set $\mathcal{A}=\{(\tilde{b}_1,\tilde{b}_2,\hat{b}_{1}, \hat{b}_{2})|\tilde{b}_1, $ $\tilde{b}_2, \hat{b}_{1}, \hat{b}_{2}\in \{0,1\}\}$). We partition $\mathcal{A}$ into $5$ subsets shown by $\mathcal{A}_g, g=1,...,5$ (see Table \ref{table3}), based on the same error probability at the relay. In fact, in each subset (group), the concentration of molecules (of each type) around the relay is the same after reaction and thus the error probability is the same. Therefore, we rewrite the error probability at the relay, given in \eqref{eq_peRPNCISI}, as
\begin{align}\label{peRPNCISI}
\mP(E_{\msR,k}|B_{1,k}=b_1,B_{2,k}=b_2)=\frac{1}{4}\sum_{g=1}^{9}f_g p_g(b_1,b_2),
\end{align}
where $f_g=\sum_{(\tilde{b}_1,\tilde{b}_2,\hat{b}_1, \hat{b}_2) \in \mathcal{A}_g}\mP\{\hat{B}_{1,k-1}^{\msT_2}=\hat{b}_1, \hat{B}_{2,k-1}^{\msT_1}=\hat{b}_2|\tilde{b}_1,\tilde{b}_2\}$, in which $\mP\{\hat{B}_{1,k-1}^{\msT_2}=\hat{b}_1, \hat{B}_{2,k-1}^{\msT_1}=\hat{b}_2|\tilde{b}_1,\tilde{b}_2\}$ is given in \eqref{eq_state}, and $$p_g(b_1,b_2)=\mP(E_{\msR,k}|B_{1,k}=b_1,B_{2,k}=b_2, (B_{1,k-1},B_{2,k-1},\hat{B}_{1,k-1}^{\msT_2},\hat{B}_{2,k-1}^{\msT_1}) \in \mathcal{A}_g).$$
% In the following, we compute $p_g(b_1,b_2)$ for each group.\\
According to \eqref{XikPNCbargashti}, the concentration of molecule type $\msM_1$ before reaction is
$c_\textrm{r} \cdot(B_{1,k}+\eta_3 \hat{B}_{2,k-1}^{\msT_1}+\eta_3B_{1,k-1})+\eta_3^2 \cdot (c_\textrm{r}\hat{B}_{2,k-2}^{\msT_1}+X_{1,k-2}\pi_1+ X_{1,k-3}\pi_3)$. The concentration of $\msM_2$ before reaction is similar. Assuming only the error in decoding the message of the previous super time slot, we have $$C_{1,k}=\max\{0,\big[B_{1,k}-B_{2,k}+\eta_3\cdot(B_{1,k-1}-\hat{B}_{1,k-1}^{\msT_2})-\eta_3\cdot(B_{2,k-1}-\hat{B}_{2,k-1}^{\msT_1})\big]c_\textrm{r}\},$$ and $C_{2,k}$ can be obtained similarly (see Table \ref{table3}).

%In group $1$, the previous messages at the transceivers are decoded without error and all interference is canceled out at the relay.
%Thus, $C_{i,k}=b_{\msR_i,k} c^{\textrm{PNC}}$. In group $2$, the previous message of $\msT_1$ is $0$ which is decoded with error as $1$ at $\msT_2$, while the previous message of $\msT_2$ is decoded without error at $\msT_1$. According to \eqref{XikPNCbargashti}, the concentration of molecule type $\msM_1$ before reaction is
%$c_\textrm{r}(B_{1,k}+\eta_3 \hat{B}_{2,k-1}^{\msT_1}+\eta_3B_{1,k-1})+\eta_3^2 (c_\textrm{r}\hat{B}_{2,k-2}^{\msT_1}+X_{1,k-2}\pi_1+ X_{1,k-3}\pi_3)$.
%The concentration of $\msM_2$ before reaction is similar. By considering the decoding error of the previous messages at the transceivers,
%we obtain $C_{1,k}=\max \{0,(B_{1,k}-B_{2,k}-\eta_3)c_\textrm{r}\}$ and $C_{2,k}=\max \{0,(B_{2,k}-B_{1,k}+\eta_3)c_\textrm{r}\}$.

%The concentration of molecules after reaction for the other groups can be obtained similarly (see Table \ref{table3}).\\
\begin{table}
\begin{center}
\caption{Concentration of molecules around the relay after reaction (PNC scheme with ISI)}
\begin{tabular}{|K{1.005cm}|K{3.2cm}|K{4.3cm}|K{4.3cm}|}
\hline
group $g$& $\mathcal{A}_g$ & $C_{1,k}$ & $C_{2,k}$\\\hline\hline
\multirow{2}{*}{$1$} & $\{(0,0,0,0), (1,0,1,0),$ & \multirow{2}{*}{$B_{\msR_1,k} c_\textrm{r}$} & \multirow{2}{*}{$B_{\msR_2,k} c_\textrm{r}$}\\
& $(0,1,0,1),(1,1,1,1),$ & &\\
& $(0,0,1,1),(1,1,0,0)\}$ & &\\\hline
\multirow{2}{*}{$2$} & $\{(0,0,1,0),(0,1,1,1),$ & \multirow{2}{*}{$\max\{0,(B_{1,k}-B_{2,k}-\eta_3) c_\textrm{r}\}$} & \multirow{2}{*}{$\max\{0,(B_{2,k}-B_{1,k}+\eta_3) c_\textrm{r}\}$}\\
& $(0,1,0,0),(1,1,1,0)\}$ & &\\\hline
\multirow{2}{*}{$3$} & $\{(0,0,0,1),(1,0,1,1),$ & \multirow{2}{*}{$\max\{0,(B_{1,k}-B_{2,k}+\eta_3) c_\textrm{r}\}$} & \multirow{2}{*}{$\max\{0,(B_{2,k}-B_{1,k}-\eta_3) c_\textrm{r}\}$}\\
& $(1,0,0,0),(1,1,0,1)\}$ & &\\\hline
%$4$ & $\{(0,0,1,1)\}$ & $B$ & $B_{\msR_2,k} c_\textrm{r}$\\\hline
%$5$ &$\{(1,1,0,0)\}$ & $B_{\msR_1,k} c_\textrm{r}$ & $\max\{0,(B_{2,k}-B_{1,k}-2\eta_3) c_\textrm{r}\}$\\\hline
$4$ & $\{(0,1,1,0)\}$ & $\max\{0,(B_{1,k}-B_{2,k}-2\eta_3) c_\textrm{r}\}$ & $\max\{0,(B_{2,k}-B_{1,k}+2\eta_3) c_\textrm{r}\}$\\\hline
$5$ & $\{(1,0,0,1)\}$ & $\max\{0,(B_{1,k}-B_{2,k}+2\eta_3) c_\textrm{r}\}$ & $\max\{0,(B_{2,k}-B_{1,k}-2\eta_3) c_\textrm{r}\}$\\\hline
\end{tabular}
\label{table3}
\end{center}
$B_{\msR_i,k}=B_{i,k}\cdot(B_{1,k}\oplus B_{2,k}), \quad i \in \{1,2\}$.
\vspace{-0.5em}
\end{table}
We assume the fixed thresholds at the relay as $\tau_1^{\msR}=\tau_2^{\msR}=0$. For group 1, since all interference is canceled out, the probability of error at the relay is equal to the no ISI case (obtained in \eqref{eq_PeRPNC_cond}). For group 2, according to Table \ref{table3}, when $(B_{1,k},B_{2,k})\in\{(0,0),(1,1)\}$, $C_{1,k}=0$ and $C_{2,k}=\eta_3 c_\textrm{r}$, and therefore, $\mP(E_{\msR_1,k})=0$ and the error probability at the relay equals to $\mP(E_{\msR_2,k})$. When $B_{1,k}=1,B_{2,k}=0$ (assuming that $\eta_3<1$), we have $C_{1,k}=(1-\eta_3)c_\textrm{r}$ and $C_{2,k}=0$, and thus, $\mP(E_{\msR_2,k})=0$. When $B_{1,k}=0,B_{2,k}=1$, we get $C_{1,k}=0$ and $C_{2,k}=(1+\eta_3) c_\textrm{r}$, and hence, $\mP(E_{\msR_1,k})=0$.
Therefore,
\begin{align}\label{PeRPNCISI11}
\nonumber
&p_2(0,0)=p_2(1,1)=1-\exp(-\eta_3 c_\textrm{r}v_\textrm{r}),\\\nonumber
&p_2(1,0)=\exp(-(1-\eta_3) c_\textrm{r}v_\textrm{r}), \quad
p_2(0,1)=\exp(-(1+\eta_3) c_\textrm{r}v_\textrm{r}).
\end{align}
\begin{table}
\begin{center}
\caption{Probability of error at the relay for the PNC scheme in the presence of ISI}
\begin{tabular}{|K{1.005cm}|K{3.2cm}|K{4.3cm}|K{4.3cm}|}
\hline
group $g$ & $p_{g}(0,0)=p_{g}(1,1)$ & $p_{g}(1,0)$ & $p_{g}(0,1)$\\\hline\hline
$1$ & $0$ & $\exp(-c_\textrm{r}v_\textrm{r})$ & $\exp(-c_\textrm{r}v_\textrm{r})$\\
$2$ & $1-\exp(-\eta_3 c_\textrm{r}v_\textrm{r})$ & $\exp(-(1-\eta_3) c_\textrm{r}v_\textrm{r})$ & $\exp(-(1+\eta_3) c_\textrm{r}v_\textrm{r})$\\
$3$ & $1-\exp(-\eta_3 c_\textrm{r}v_\textrm{r})$ & $\exp(-(1+\eta_3) c_\textrm{r}v_\textrm{r})$ & $\exp(-(1-\eta_3) c_\textrm{r}v_\textrm{r})$ \\
$4$ & $1-\exp(-2\eta_3 c_\textrm{r}v_\textrm{r})$ &
$\exp(-|1-2\eta_3|c_\textrm{r}v_\textrm{r})$ & $\exp(-(1+2\eta_3) c_\textrm{r}v_\textrm{r})$\\
$5$ & $1-\exp(-2\eta_3 c_\textrm{r}v_\textrm{r})$ &
$\exp(-(1+2\eta_3) c_\textrm{r}v_\textrm{r})$ &
$\exp(-|1-2\eta_3| c_\textrm{r}v_\textrm{r})$\\
\hline
\end{tabular}
\label{table4}
\end{center}
\vspace{-2.5em}
\end{table}

\end{document}